\documentclass[%
reprint,
amsmath,amssymb,
aps,
pra,
]{revtex4-2}

\usepackage{graphicx}% Include figure files
\usepackage{dcolumn}% Align table columns on decimal point
\usepackage{bm}% bold math
\begin{document}

%\preprint{APS/123-QED}

\title{Dirac Equation for Photons: Origin of Polarisation}% Force line breaks with \\

\author{Shinichi Saito}
 \email{shinichi.saito.qt@hitachi.com}
\affiliation{Center for Exploratory Research Laboratory, Research \& Development Group, Hitachi, Ltd. Tokyo 185-8601, Japan.}%Lines break automatically or can be forced with \\
%\author{Isao Tomita}%
%\affiliation{% Department of Electrical and Computer Engineering, National Institute of Technology, Gifu College, 2236-2 Kamimakuwa, Motosu, Gifu 501-0495, Japan.}%

\date{\today}% It is always \today, today, %  but any date may be explicitly specified

\begin{abstract}
Spin is a fundamental degree of freedom, whose existence was proven by Dirac for an electron by imposing the relativity to quantum mechanics, leading to the triumph to derive the Dirac equation.
Spin of a photon should be linked to polarisation, however, the similar argument for an electron was not applicable to Maxwell equations, which are already Lorentz invariant.
Therefore, the origin of polarisation and its relationship with spin are not completely elucidated, yet.
Here, we discuss propagation of coherent rays of photons in a graded-index optical fibre, which can be solved exactly using the Laguerre-Gauss or Hermite-Gauss modes in a cylindrical or a Cartesian coordinate.
We found that the energy spectrum is massive with the effective mass as a function of the confinement and orbital angular momentum.
The propagation is described by the one-dimensional ($1D$) non-relativistic Schr\"odinger equation, which is equivalent to the $2D$ space-time Klein-Gordon equation by a unitary transformation.
The probabilistic interpretation and the conservation law require the factorisation of the Klein-Gordon equation, leading to the $2D$ Dirac equation with spin.
We applied the Bardeen-Cooper-Schrieffer (BCS)-Bogoliubov theory of superconductivity to a coherent ray from a laser and identified a radiative Nambu-Anderson-Higgs-Goldstone mode for recovering the broken symmetry.
The spin expectation value of a photon corresponds to the polarisation state in the Poincar\'e sphere, which is characterised by fixed phases after the onset of lasing due to the broken $SU(2)$ symmetry, and it is shown that its azimuthal angle is coming from the phase of the energy gap.
\end{abstract}
%Max 600 characters in PRL, 500 words for PRA & PRB

\maketitle
%\tableofcontents

%\begin{figure}[h]
%\begin{center}
%\includegraphics[width=8.6cm]{Fig1.pdf}
%\caption{
 %New $3D$ carbon allotrope, Hopfene.
%(a) Crystal structure. 
%}
%\end{center}
%\end{figure}

\section{Introduction}
Dirac elucidated the origin of spin for an electron \cite{Dirac28,Dirac30} by unifying quantum mechanics \cite{Plank00,Einstein05,Bohr13} with the theory of relativity \cite{Einstein05b}, which led to the discovery of the Dirac equation \cite{Dirac30, Baym69,Sakurai14,Sakurai67}.
In order to accept the probabilistic interpretation of the wavefunction together with the conservation law, it was inevitable to factorise the Klein-Gordon equation with non-trivial spin commutation relationship in a matrix form, called as a spinor representation \cite{Dirac30, Sakurai67}.
Therefore, the Lorentz invariance in the fundamental Dirac equation was a major principle behind the existence of spin for an electron.

However, the corresponding argument for a photon is not straightforward, since the Maxwell equations are already invariant under the Lorentz transformation \cite{Jackson99}.
The classical electro-magnetic waves, derived from the Maxwell equations, already describe the propagation of light \cite{Jackson99,Yariv97}, even without introducing quantum mechanics. 
The polarisation of light is naturally understood by the transverse nature of electro-magnetic radiation \cite{Goldstein11,Gil16,Pedrotti07,Hecht17}, but it is less obvious how the quantum mechanical nature of spin for a photon is required to understand the polarisation \cite{Lehner14}.

Historically, at the early stage of the quantum mechanics, there are several attempts \cite{Landau30,Oppenheimer31,Laporte31,Darwin32} to derive a photonic counter part of the Dirac equation.
Unfortunately, these works are not successful for photons, compared with electrons.
More recently, spin angular momentum for photons is revisited in close relationship with optical orbital angular momentum \cite{Allen92,Enk94,Leader14,Barnett16,Yariv97,Jackson99,Grynberg10,Bliokh15,Enk94,Leader14,Barnett16,Chen08,Ji10} in order to understand the fundamental quantum mechanical nature of photons \cite{Lehner14}.
So far, it is widely considered that the optical angular momentum is well-defined, however, it is impossible to split into spin angular momentum and orbital angular momentum \cite{Allen92,Enk94,Leader14,Barnett16,Yariv97,Jackson99,Grynberg10,Bliokh15} in a gauge invariant way \cite{Enk94,Leader14,Barnett16,Chen08,Ji10}.
Therefore, the spin degree of freedom for a photon is not completely well-defined in a quantum-mechanical way.

We have recently considered {\it what is spin of a photon} by using a standard quantum many-body field theory \cite{Saito20a}.
We assumed that a coherent ray of photons emitted from a laser source, described by a coherent state with macroscopic number of photons in a waveguide, and showed that the quantum-mechanical spin operators, $\hat{\bf S}$, are derived by using a $SU(2)$ Lie algebra \cite{Saito20a}.
Contrary to the popular view that the coherent ray is classical, we showed that the polarisation state is described by the quantum-mechanical average of spin operators, $\langle \hat{\bf S} \rangle$, because of the Bose-Einstein condensation of photons \cite{Saito20a}.
The phase and the amplitude of the macroscopic wavefunction can be easily controlled by passive optical components such as a phase-shifter and a rotator \cite{Yariv97,Jackson99,Goldstein11,Gil16,Pedrotti07,Hecht17}, while the $SU(2)$ group theory is a powerful tool to describe the spin state \cite{Jones41,Payne52,Max99,Yariv97,Baym69,Sakurai14,Yariv97,Collett70,Luis02,Luis07,Bjork10,Castillo11,Sotto18,Sotto18b,Sotto19,Saito20a}.
It was shown that the spin expectation value, $\langle \hat{\bf S} \rangle$, is exactly the same as Stokes parameters $(S_1,S_2,S_3)$ in Poincar\'e sphere \cite{Stokes51,Poincare92,Saito20a}.
Therefore, we believe that spin of a photon is well-defined quantum-mechanically and spin is an inherent property of a photon as an elementary particle.

We have also analysed the commutation relationship of orbital angular momentum operators for a photon  \cite{Saito20b}, described by a Laguerre-Gauss mode \cite{Allen92,Yariv97} .
We showed that the ladder operators to increment and decrement the orbital angular momentum along the direction of the propagation in the unit of the Dirac constant, $\hbar=h/(2\pi)$, where $h$ is the Plank constant  \cite{Saito20b}.
We have also shown the validity of the quantum commutation relationship against the Laguerre-Gauss mode, and the Laguerre-Gauss mode is labelled by a definite integer quantum number ($m$) to describe the magnetic orbital angular momentum of $\hbar m$.
The Laguerre-Gauss mode is not an eigenstate for the magnitude of the orbital angular momentum, $\hat{\bf l} \cdot \hat{\bf l}$, but nevertheless, the quantum expectation value, $\langle \hat{\bf l} \cdot \hat{\bf l} \rangle$, was analytically calculated, which contains both intrinsic and extrinsic contributions \cite{Saito20b}.
Based on these results, it was suggested that the orbital angular momentum is also a well-defined physical observable for a photon.

Then, we have revisited the issue of the proper splitting of the total photonic orbital angular momentum into spin and orbital angular momentum, which was considered to be impossible to achieve  \cite{Allen92,Enk94,Leader14,Barnett16,Yariv97,Jackson99,Grynberg10,Bliokh15,Enk94,Leader14,Barnett16,Chen08,Ji10} 
without imposing special boundary conditions \cite{Chen08,Ji10,Leader14} in a proper gauge invariant way.
We considered the propagation of the Laguerre-Gauss mode in a GRaded-INdex (GRIN) fibre \cite{Kawakami68} and found that the splitting is indeed possible  \cite{Saito20c}.
One of the most important part of this derivation was to assume the finite mode size of the wave rather than assuming a simple plane wave expansions.
Historically, quantum mechanics was developed by the consideration of the black body radiation \cite{Plank00,Einstein05,Bohr13,Baym69}, for which the plane-wave expansion was appropriate to consider all possible modes in a black body.
The field theoretic treatments using creation and annihilation operators, known as the second quantisation techniques, were developed based on the plane wave expansion \cite{Dirac30,Sakurai14,Sakurai67}.
However, the plane waves with continuous spectra were not suitable for describing the discrete confined modes with proper symmetries reflecting the profile of the refractive index in a waveguide.
The mode profile is essentially described by a wavefunction of a photon, which is obtained as a solution of the Helmholtz equation \cite{Saito20c}. 
Consequently, there exists the small longitudinal electric field, coming from the finite shape of the envelope wavefunction, which ensures the proper splitting between spin and orbital angular momentum from the total angular momentum.
From the classical correspondence of the angular momentum density operator, $\hat{\bf m}={\bf r} \times \hat{\bf p}$, between the momentum density operator, $\hat{\bf p}$, and the coordinate, ${\bf r}$, we obtained the orbital angular momentum operator for spin and orbital angular momentum along the direction of propagation  \cite{Saito20c}.
The projected components of spin and orbital angular momentum operators along the direction of propagation are helicity operators.
For the transverse components of the angular momentum, we accepted the principle of the quantum-mechanical rotational symmetries for spin and orbital angular momentum using the $SU(2)$ Lie algebra, which led to the expressions for all components of spin and orbital angular momentum operators  \cite{Saito20c}.
The spin expectation value becomes Stokes parameters in the Poincar\'e sphere, as expected, and the expectation value of orbital angular momentum is also described by similar parameters in the higher-order Poincar\'e sphere \cite{Padgett99,Milione11,Liu17,Erhard18,Saito20c}.
Based on these analyses \cite{Saito20a,Saito20b,Saito20c}, we are approaching to understand some quantum-mechanical features of spin and orbital angular momentum for coherent photons.
However, we have not established the intrinsic quantum-mechanical origin of spin for a photon, as compared with the level of understanding of spin for an electron using the Dirac equation \cite{Dirac28,Dirac30}.

In this paper, we discuss why spin is inevitable to guarantee the quantum-mechanical probabilistic interpretation, which is based on exactly the same principle for the existence of spin for an electron \cite{Dirac28,Dirac30}.
We also discuss how spin of a photon is linked to the origin of polarisation, as a manifestation of macroscopic realisation of a quantum state of light.
In order to make the argument specific, we restrict our analysis on propagation of a coherent ray of photons in a GRIN fibre \cite{Kawakami68,Yariv97,Saito20a,Saito20b,Saito20c}, but the extension to the other waveguide will be straightforward.
The advantage to employ the GRIN fibre is availability of the exact solution of the Helmholtz equation, and therefore we can theoretically treat exactly. 
We obtained two-dimensional ($2D$) space-time Dirac equation to describe the propagation of photons along the GRIN fibre, and found that spin is naturally derived to account for the polarisation state.
The rotational invariance of a polarisation state in a waveguide with a cylindrical symmetry is a fundamental principle for the quantum nature of spin for photons.
For the coherent photons from a laser source, the many-body state is described by a coherent state \cite{Grynberg10,Fox06,Parker05}, which is a superposition state made of states with different number of photons.
Thanks to this uncertainty of number of photons, the phase is coherently fixed and a macroscopic number of photons are occupying the same state, which can be regarded as Bose-Einstein condensation \cite{Grynberg10,Fox06,Parker05,Nagaosa99,Wen04,Altland10,Saito20a,Saito20b,Saito20c}.
We introduce the concept of the broken symmetry \cite{Nambu59}, which is established for the Bardeen-Cooper-Schrieffer (BCS) theory of superconductivity \cite{Bardeen57,Anderson58,Schrieffer71,Abrikosov75,Nagaosa99,Wen04,Altland10}, to photons.
In case of photons, the phase coherence is achieved by the onset of lasing, and we show that the energy gap, $|\Delta|$, is indeed opening up due to the confinement of photons in a waveguide.
It is also well-established that there exists a boson to recover the symmetry of the system, whose energy should vanish in the long wavelength limit, known as a Nambu-Anderson-Higgs-Goldstone mode \cite{Nambu59,Anderson58,Goldstone62,Higgs64}.
In our case, this corresponds to a radiative mode, escaping from the waveguide.
By applying the Bogoljubov transformation \cite{Bogoljubov58,Schrieffer71,Abrikosov75}, we show that the phase, $\phi$, of the order parameter, $\Delta=|\Delta|{\rm e}^{i \phi}$, corresponds to the azimuthal angle of the polarisation state in the Poincar\'e sphere.

\section{Principles}
\subsection{Helmholtz equation}
We discuss propagation of a photon in a material with refractive index profile of $\epsilon=\epsilon({\bf r})$ as a function of the position, ${\bf r}=(x,y,z)$.
It is interesting to be aware that the wavefunction of a photon and the fundamental equation to describe the wavefunction are not frequently discussed.
However, a photon is an elementary particle, such that the probability to find a photon at a certain position (${\bf r}$) is described by a wavefunction, $\Psi ({\bf r})$, which satisfies the Helmholtz equation \cite{Jackson99,Yariv97,Saito20a,Saito20b,Saito20c}
\begin{eqnarray}
\nabla^2 
\Psi({\bf r})
=
\mu_0 \epsilon ({\bf r})
\frac{\partial^2}{\partial t^2}
\Psi({\bf r}),
\end{eqnarray}
where $\nabla=(\partial_x, \partial_y,\partial_z)$, $t$ is time, $\mu_0$ is the vacuum permeability, and $ \epsilon = \epsilon ({\bf r})$ is the profile of the dielectric constant of a material.
The values of permeability for most of materials used in photonics are almost the same as that in vacuum.
In the limit of a vacuum, $\epsilon \rightarrow \epsilon_0$, where $\epsilon_0$ is the vacuum permittivity, the Helmholtz equation becomes
\begin{eqnarray}
\nabla^2 
\Psi({\bf r})
=
\frac{1}{c^2}
\frac{\partial^2}{\partial t^2}
\Psi({\bf r}),
\end{eqnarray}
where $c=1/\sqrt{\mu_0 \epsilon_0}$ is the velocity of light in a vacuum.
In a vacuum, the dispersion relationship between the angular frequency $\omega_0$ and the wavenumber $k_0=2 \pi / \lambda$ for the wavelength of $\lambda$ becomes linear $\omega_0 = ck_0$, and the wavefunction just becomes a simple plane wave, $\Psi({\bf r})={\rm e}^{i(k_0 z - \omega_0 t)}$, assuming the direction of propagation is along $z$.
The linear dispersion means that a photon does not possess rest mass as an elementary particle \cite{Sakurai67}.
In a vacuum, there is no confinement of a photon, and the mode profile is spreading to the entire system.
In reality, the mode profile could be adjusted by optical lens \cite{Yariv97} with a diffraction limited mode profile.

On the other hand, in a material with the refractive index profile of $n({\bf r})=\sqrt{\epsilon({\bf r})/\epsilon_0}$, a photon has a tendency to propagate in a region where the refractive index is large, to minimise the optical path length, following the Fermat's principle and the Eikonal equation \cite{Yariv97,Saito20a,Saito20b,Saito20c}.
The dispersion relationship $\omega=\omega(k)$ for the wavenumber, $k$, in a material must be obtained by solving the Hemholtz equation.
Depending on $n({\bf r})$, we expect various non-trivial solutions. 
For example, if $n({\bf r})$ is a periodic function against ${\bf r}$, the dispersion relationship has a bad gap structure, which is very similar to the electronic band structure for electrons in a material, and the corresponding photonic strucutre is called as a photonic crystal \cite{Joannopoulos08,Sotto18,Sotto18b,Sotto19}.

\subsection{Quantisation}
We assume that the direction of propagation is along $z$ and the fibre has a cylindrical symmetry.
In a GRIN fibre, the refractive index, $n({\bf r})=n(r)$, satisfies 
\begin{eqnarray}
n(r)^2
=
n_0^2
\left(
1-g^2 r^2
\right),
\end{eqnarray}
where $g$ is the parameter to account for the quadratic graded-index profile with the dimension of the inverse of the length, and the radius is defined as $r=\sqrt{x^2+y^2}$ in a cylindrical coordinate of $(r,\phi,z)$.
We assume that the solution is decoupled between the direction of the propagation and the mode profile, such that the solution is given by a product form, 
\begin{eqnarray}
\Psi({\bf r})=\psi(x,y) {\rm e}^{i(k z - \omega_0 t)}.
\end{eqnarray}
By inserting this expression into the Helmholtz equation, we obtain
\begin{eqnarray}
\left(
  \left(
\partial_x^2 
+
\partial_y^2 
-\frac{\omega_0^2 g^2}{v_0^2}r^2
  \right)
+
  \left(
-k^2
+
\frac{\omega_0^2}{v_0^2}
  \right)
\right)
\psi (x,y)
=0,
\nonumber \\
\end{eqnarray}
where $v_0=c/n_0$.
The dispersion relationship is not just a sum along the perpendicular directions, and it must be solved by a quadratic equation, as we shall show below.
We will solve this equation with the help of special functions.

In a Cartesian coordinate, the solution \cite{Yariv97,Saito20c} is given by
\begin{eqnarray}
\psi(x,y)
&=
H_l
\left(
  \sqrt{2}\frac{x}{w_0}
\right)
H_m
\left(
  \sqrt{2}\frac{y}{w_0}
\right)
{\rm  e}^{-\frac{r^2}{w_0^2}},
\end{eqnarray}
where 
$w_0=
\sqrt{
  2/(gk_{n_0})
}$
with 
$k_{n_0}= 2 \pi /  \lambda_{n_0}
=2 \pi n_0/ \lambda
=
k_0 n_0
=\omega_0 n_0/c
=
\omega_0/v_0
$,
where $H_l$ and $H_m$ are Hermite polynomials and $l=0,1,2,\cdots$ and $m=0,1,2,\cdots$ are quantum numbers to account for the number of nodes along $x$ and $y$.
This Hermite-Gaauss solution satisfies
\begin{eqnarray}
\left(
\partial_x^2 
+
\partial_y^2 
-\frac{\omega_0^2 g^2}{v_0^2}r^2
\right)
\psi
=
-2 \frac{g}{v_0} (l+m+1) \omega_0 \psi.
\end{eqnarray}

The more natural solution is given by a cylindrical coordinate as, 
\begin{eqnarray}
\psi(r,\phi)
=&
\left(
\frac{\sqrt{2}r}{w_0}
\right)^{|m|}
L_n^{|m|} 
\left(
2
\left(
  \frac{r}{w_0}
\right)^2 
\right) 
{\rm  e}^{-\frac{r^2}{w_0^2}}
{\rm  e}^{im\phi},
\end{eqnarray}
where $L_n^m$ is the associated Laguerre polynomial with the radial quantum number of $n=0, 1, 2 \cdots$ and the magnetic angular momentum of $m=0,\pm 1, \pm 2, \cdots$.
This Laguerre-Gauss solution satisfies
\begin{eqnarray}
\left(
\partial_r^2 
+
\frac{1}{r}
\partial_r 
+
\frac{1}{r^2}
\partial_{\phi}^2 
-\frac{\omega_0^2 g^2}{v_0^2}r^2
\right)
\psi
=
-2 \frac{g}{v_0} (2n+|m|+1) \omega_0 \psi.
\nonumber \\
\end{eqnarray}
By assigning $l\rightarrow 2n$ and $m \rightarrow |m|$, both Hermite-Gauss and Laguerre-Gauss solutions satisfy the quadratic equation for the dispersion relationship
\begin{eqnarray}
\omega_0^2  -2 \delta w_0 (2n+|m|+1) \omega_0 - v_0^2 k^2 =0,
\end{eqnarray}
where the shift of the angular frequency is defined as $\delta w_0=v_0 g$.
By multiplying $\hbar^2$, this is also rewritten as 
\begin{eqnarray}
(\hbar \omega_0)^2  -2 (\hbar \delta  w_0) (2n+|m|+1) (\hbar \omega_0) - (\hbar v_0 k)^2 =0.
\nonumber \\
\end{eqnarray}
Here, we assume quantisation conditions for the energy, $E_0$, and the momentum, $p$, as \cite{Plank00,Einstein05,Bohr13,Tomonaga62,Tomonaga66,Nauenberg16,Baym69,Sakurai14},
\begin{eqnarray}
E_0 &=&\hbar \omega_0 \\
p &=& \hbar k.
\end{eqnarray}
We also define an energy gap as
\begin{eqnarray}
\Delta
&=&\hbar \delta  w_0 (2n+|m|+1)  \\
&=&m^{*} v_0^2,
\end{eqnarray}
which is similar to the famous Einstein's equation $E=m c^2$ \cite{Einstein05c}.
In fact, $m^{*}$ is the effective mass of the photon in the GRIN fibre, which is not zero in a dispersive GRIN fibre.
The effective mass increases upon increasing $n$ and $|m|$ and strengthening the optical confinement by increasing $g$.
Please note that $\Delta$ is finite even at $n=0$ and $m=0$, which means that the confinement in the fibre makes a photon massive.
By inserting these expressions, the quadratic dispersion equation becomes
\begin{eqnarray}
E_0^2  -2 \Delta E_0 - (v_0 p)^2 =0.
\end{eqnarray}
This equation formally gives two solutions: one for a particle
\begin{eqnarray}
E_0^{+}=
\Delta
+
\sqrt{\Delta^2+ (v_0 p)^2}
\end{eqnarray}
with positive energy, and the other is for an anti-particle
\begin{eqnarray}
E_0^{-}
=
\Delta
-
\sqrt{\Delta^2+ (v_0 p)^2}
\end{eqnarray}
with negative energy.
It is natural to expect the positive particle solution for assigning to the standard guided optical mode, while the existence of the gapless mode would be linked to the Nambu-Anderson-Higgs-Goldstone mode \cite{Nambu59,Anderson58,Goldstone62,Higgs64}, as we shall see.

\subsection{$1D$ Schr\"odinger equation for a photon}
In quantum mechanics, energy-time and momentum-position have dualities governed by the uncertainty principle \cite{Dirac30,Tomonaga62,Tomonaga66,Baym69,Sakurai14}.
Then, it is natural to expect
\begin{eqnarray}
E_0
&=&i \hbar \partial_t \\
p
&=&
\frac{\hbar}{i} \partial_z ,
\end{eqnarray}
and consequently, the quadratic dispersion equation becomes
\begin{eqnarray}
- i \hbar \partial_t {\rm e}^{i k z - i \omega_0 t}
=
-
\frac{\hbar^2}{2 m^{*}}
\left[
  \partial_z^2-\frac{1}{v_0^2} \partial_t^2
\right]
{\rm e}^{i k z - i \omega_0 t}.
\end{eqnarray}
If we define the space-time ($2D$) d'Alembertian as 
\begin{eqnarray}
\Box_2
=  
\partial_z^2-\frac{1}{v_0^2} \partial_t^2,
\end{eqnarray}
we obtain 
\begin{eqnarray}
- i \hbar \partial_t \psi_z
=
-
\frac{\hbar^2}{2 m^{*}}
\Box_2 
\psi_z,
\end{eqnarray}
where the wavefunction along $z$ is given by a simple plane wave form as
\begin{eqnarray}
\psi_z = {\rm e}^{i k z - i \omega_0 t}.
\end{eqnarray}
This is essentially the same as the same as the non-relativistic $1D$ Schr\"odinger equation, although the sign in front of the time derivative is opposite.
Please note that it is not required to impose Lorentz invariance in a material, because the velocity of light is not $c$ any more and it depends on the choice of the frame.
Consequently the obtained non-relativistic Schr\"odinger equation is not symmetric between space and time.
In this paper, we are discussing in the rest frame,  and the application of the special theory of relativity for the moving frame will be discussed elsewhere.

In the limit of the vanishing effective mass, $ m^{*} \rightarrow 0$, this becomes the correct form
\begin{eqnarray}
\Box_2 
\psi_z=0,
\end{eqnarray}
which give the massless dispersion of $\omega_0=v_0 k$ with the renormalised velocity of $v_0=c/n_0$ due to the virtual generations and absorptions of electron-hole pairs, as formulated by Feynman diagrams \cite{Fetter03}.

\subsection{Nambu-Anderson-Higgs-Goldstone mode}
We are considering the coherent ray from a laser source propagating in a GRIN fibre.
The propagation is only possible when the phase coherence is achieved due to the pumping above the lasing threshold, otherwise the phase is randomised and the order parameter, $\Delta$, becomes zero upon averaging over $t$.
The situation is very similar to the onset of superconductivity, for which the quasi-particle spectrum becomes massive\cite{Bardeen57,Nambu59} and a gapless Nambu-Anderson-Higgs-Goldstone mode appears to recover the symmetry of the broken symmetry \cite{Nambu59,Anderson58,Goldstone62,Higgs64}.

In the previous subsection, we obtained the anti-particle gapless solution, such that we discuss the similarity.
While it is impossible to accept the realisation of the propagation with negative energy unless the time is evolving in a reverse way, it is possible to consider a leaky mode from the fibre by assuming a diverging solution 
\begin{eqnarray}
{\rm  e}^{-\frac{r^2}{w_0^2}}
\rightarrow \ 
{\rm  e}^{+\frac{r^2}{w_0^2}},
\end{eqnarray}
which corresponds to 
\begin{eqnarray}
\Delta \rightarrow \ -\Delta.
\end{eqnarray}
Then, the $1D$ Schr\"odinger equation becomes
\begin{eqnarray}
i \hbar \partial_t \psi_z^{-}
=
-
\frac{\hbar^2}{2 m^{*}}
\Box_2 
\psi_z^{-},
\end{eqnarray}
where $\psi_z^{-}$ is the wavefunction of the leaky mode.
The eigenvalues become
\begin{eqnarray}
E_0^{-\pm}
=
-\Delta
\pm
\sqrt{\Delta^2+ (v_0 p)^2},
\end{eqnarray}
among which one of the solution provides a positive solution
\begin{eqnarray}
E_0^{-+}
&=&
-\Delta
+
\sqrt{\Delta^2+ (v_0 p)^2}
\\
&\approx&
\frac{(v_0 p)^2}{2\Delta}
=
\frac{p^2}{2m^{*}}.
\end{eqnarray}
This dispersion is gapless, which means the arbitrary small energy is required to recover the symmetry.
In the case of propagation in a fibre, the symmetry is broken, because of the confinement in the fibre.
The existence of the gapless mode guarantees that the existence of a mode to recover the confinement by escaping from the fibre.
In our case, the so-called Nambu-Goldstone boson is also made of a photon with the leaking branch.
In reality, the leaky mode is completely different in energy and the momentum, such that the guided mode is not affected by the leaky mode at all.
It is just theoretically interesting to consider the fundamental principle behind the photon propagation in the GRIN fibre.

If we use the same notation to the leaky mode, the wavefunction of the guided mode, $\psi_z^{+}$, must satisfy the time-reversal $1D$ Schr\"odinger equation
\begin{eqnarray}
- i \hbar \partial_t \psi_z^{+}
=
-
\frac{\hbar^2}{2 m^{*}}
\Box_2 
\psi_z^{+},
\end{eqnarray}
which will give 2 formal solutions 
\begin{eqnarray}
E_0^{+\pm}
=
\Delta
\pm
\sqrt{\Delta^2+ (v_0 p)^2}
\end{eqnarray}
for the standard time evolution and the reverse time evolution.
All solutions are summarised in Table \ref{Table-I}.

\begin{table}[h]
\caption{\label{Table-I}
Dispersion relationships for a graded-index (GRIN) fibre.
The guided mode has a gapful dispersion for a standard time evolution, while the leaky mode is gapless.
The reverse time is only meaningful theoretically, or it is valid for describing the light propagation to the opposite direction, $k<0$.
}
\begin{ruledtabular}
\begin{tabular}{ccc}
&
Standard time & Reverse time \\
\colrule
$\psi_z^{+}$  (Guided)
&
$E_0^{++}>0$ (Gapful)
& 
$E_0^{+-}<0$ (Gapless)
\\
$\psi_z^{-}$ (Leaky)
&
$E_0^{-+}>0$ (Gapless)
& 
$E_0^{--}<0$ (Gapful) 
\end{tabular}
\end{ruledtabular}
\end{table}

\subsection{Symmetries}
Here, we focus on the symmetries of the possible solutions in a GRIN fibre.
We go back to the original Helmholtz equation
\begin{eqnarray}
\left[
\nabla^2
-
\frac{1}{v_0^2} 
\left(
1-g^2 r^2
\right)
\partial_t^2
\right]
\Psi({\bf r})
=0,
\end{eqnarray}
and discuss why several solutions appeared.
In deriving the dispersion relationship, we realise
\begin{eqnarray}
\partial_z^2
&=&
\left( ik \right)^2
=-k^2 \\
\partial_t^2
&=&
\left( i\omega \right)^2
=-\omega^2,
\end{eqnarray}
and thus the solutions are available in dependent of the signs of $k$ and $\omega_0$.
Therefore, the wavefunction along the direction of the propagation is expressed as
\begin{eqnarray}
\psi_z = {\rm e}^{\pm i k z \pm i \omega_0 t},
\end{eqnarray}
where 4 possible solutions are available depending on the choice of signs.
If we consider standard time evolution of $t>0$, $k>0$, and $\omega_0>0$, the direction of propagation is along the positive $z$ direction ($+z$) for the solutions ${\rm e}^{i k z - i \omega_0 t}$ and ${\rm e}^{-i k z + i \omega_0 t}$, and it is negative ($-z$) for ${\rm e}^{- i k z - i \omega_0 t}$ and ${\rm e}^{i k z + i \omega_0 t}$.
Alternatively, if we allow $k$ to be both negative and positive, we can always use ${\rm e}^{i k z - i \omega_0 t}$, while we should consider 2 branches for each mode: 
Gapful modes (Fig. 1 (b)) have dispersions $\hbar \omega_0=\hbar \omega_0(k)=\Delta+\sqrt{\Delta^2+ (v_0 p)^2}$ and $\hbar \omega_0=- \hbar \omega_0(k)$. 
Gapless modes (Fig. 1 (c)) have dispersions  $\hbar \omega_0=\hbar \omega_0(k)=\Delta-\sqrt{\Delta^2+ (v_0 p)^2}$ and $\hbar \omega_0=- \hbar \omega_0(k)$.

\begin{figure}[h]
\begin{center}
\includegraphics[width=5cm]{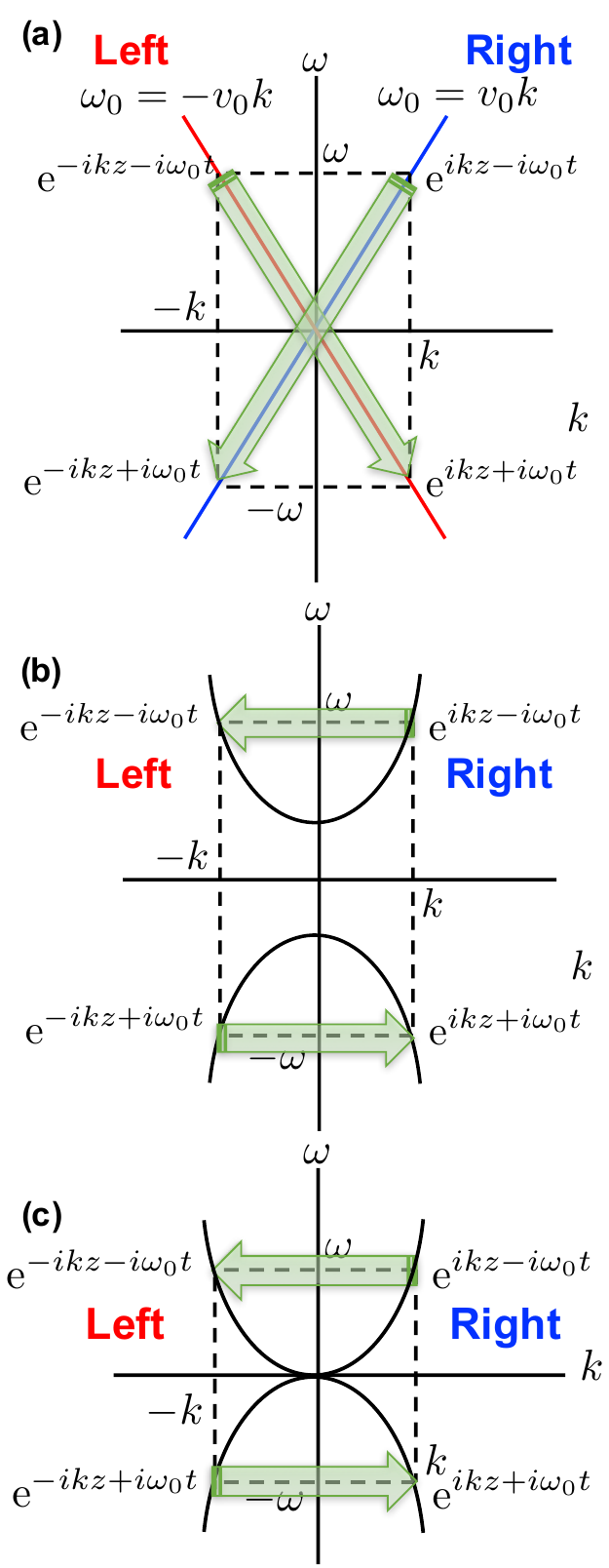}
\caption{
Schematic dispersion relationships for a graded-index (GRIN) fibre. 
(a) In the free confinement limit ($g=0$), the dispersions are just linear.
2 branches exist: one for a right propagating mode and the other is for a left propagation mode. 
There is no branch exchange upon the parity operation ($-k \leftrightarrow k$) in the momentum space.
The original right propagating state is still describing for the right mode upon the parity operation, and {\it vice versa}.
(b) In the GRIN fibre ($g\neq 0$), the guided mode has be opening up an energy gap and the spectra are massive.
The parity operation makes the branch exchange from the right to the left and {\it vice versa}, while keeping the sign of the energy unchanged. The existence of the complex conjugate solution guarantees the observable nature of electro-magnetic fields.
(c) For the radiative modes from the GRIN fibre ($g\neq 0$), the energy spectra are gapless. Still, the branch exchange is expected upon the parity operation.
}
\end{center}
\end{figure}

In the absence of the confinement ($g=0$), the massless linear dispersion, $\omega_0=\pm v_0 k$, is expected and the re are 2 branches: one for a right moving photon and the other is for a left moving photon.
The phase velocity is simply obtained 
\begin{eqnarray}
v_{\rm p}=\frac{c}{n(\omega)}
=\frac{\omega_0 (k)}{k},
\end{eqnarray}
while the group velocity \cite{Jackson99,Yariv97} is calculated by the derivative
\begin{eqnarray}
v_{\rm g}=
\left(
  \frac{d\omega_0}{dk}
\right)_{\omega_0}.
\end{eqnarray}
There is no difference in sign between $v_{\rm p}$ and $v_{\rm g}$, but $v_{\rm g}$ is always smaller than $v_{\rm p}$ in the rest frame.
For the complete linear dispersion at $g=0$, these velocities are the same, $v_{\rm p} = v_{\rm g}$.

We consider a parity exchange operation in the momentum space,
\begin{eqnarray}
\hat{\mathcal P}_{k}
k
=
-k,
\end{eqnarray}
which converts the direction of the propagation.
For $g=0$, the branch of the dispersion is not changed (Fig. 1 (a)), and thus we find
\begin{eqnarray}
\hat{\mathcal P}_{k}
\omega_0
=
-
\omega_0,
\end{eqnarray}
and 
\begin{eqnarray}
\hat{\mathcal P}_{k}
v_p
=
v_p.
\end{eqnarray}
We also consider the parity operation to the chirality, $\chi_z$, which is the projection of the spin expectation value ${\bf S}=\langle (\hat{\bf S})\rangle=(\langle \hat{S}_x \rangle,\langle \hat{S}_y \rangle,\langle \hat{S}_z\rangle)$ along the direction of the propagation \cite{Sakurai67,Barnett12,Saito20c}.
The spin state changes the sign
\begin{eqnarray}
\hat{\mathcal P}_{k}
S_z
=
-
S_z,
\end{eqnarray}
because of the change of the sign of the frequency, which changes the left circulation to the right circulation and {\it vice versa}.
On the other hand, the direction of the propagation is the same, such that the chirality is the same sign with $S_z$ as
\begin{eqnarray}
\hat{\mathcal P}_{k}
\chi_z
=
-
\chi_z.
\end{eqnarray}

In the GRIN fibre $g\neq 0$, the guided modes have an energy gap, $\Delta$, (Fig. 1 (b)), which can be regarded as a coupling between left and right moving modes, as we shall we in the next section.
This is very similar to superconductivity, explained by the theory to form a Cooper pair between electrons with the opposite momentum \cite{Bardeen57,Schrieffer71}.
It also has a similar mathematical feature with the Tomonaga-Luttinger liquid for $1D$ electron systems \cite{Tomonaga50,Luttinger63,Giamarchi04}.
Consequently, the branch exchange is expected upon the parity operation, 
\begin{eqnarray}
\hat{\mathcal P}_{k}k=-k, 
\end{eqnarray}
such that the energy will not change the sign
\begin{eqnarray}
\hat{\mathcal P}_{k}
\omega_0
=
\omega_0,
\end{eqnarray}
which is remarkable difference from a free photon.
Therefore, the spin is pointing to the same direction
\begin{eqnarray}
\hat{\mathcal P}_{k}
S_z
=
S_z
\end{eqnarray}
On the other hand, the direction of the propagation is flipped
\begin{eqnarray}
\hat{\mathcal P}_{k}
v_p
=
-
v_p,
\end{eqnarray}
as before, such that the chirality is also flipped
\begin{eqnarray}
\hat{\mathcal P}_{k}
\chi_z
=
-
\chi_z.
\end{eqnarray}

For the gapless mode (Fig. 1(c)), we expect the radiative solution, as we discussed in the previous section.
We also expect the same parity operations with the case for the gapped mode, because the spectra depends quadratic on $p=\hbar k$.
The summary of the parity operation is given in Table \ref{Table-II}.

\begin{table}[h]
\caption{\label{Table-II}
Parity operation for modes in a free space and a graded-index (GRIN) fibre.
}
\begin{ruledtabular}
\begin{tabular}{ccc}
&
Free space & GRIN fibre \\
\colrule
Momentum
&
$\hat{\mathcal P}_{k} k = -k$ 
& 
$\hat{\mathcal P}_{k} k = -k$ 
\\
Energy
&
$\hat{\mathcal P}_{k} \omega_0 = -\omega_0$ 
& 
$\hat{\mathcal P}_{k} \omega_0 = \omega_0$ 
\\
Velocity
&
$\hat{\mathcal P}_{k} v_p =  v_p$ 
& 
$\hat{\mathcal P}_{k} v_p = - v_p$ 
\\
Chirality
&
$\hat{\mathcal P}_{k} \chi_z = -\chi_z$ 
& 
$\hat{\mathcal P}_{k} \chi_z = -\chi_z$ 
\\
Spin
&
$\hat{\mathcal P}_{k} S_z = -S_z$ 
& 
$\hat{\mathcal P}_{k} S_z = S_z$ 
\end{tabular}
\end{ruledtabular}
\end{table}

If we consider both guided and radiative modes, we have 8 possible solutions for the Helmholtz equation of the GRIN fibre, among which 4 solutions are physical feasible (Table \ref{Table-III}).
There are 2 possibility for the spectra: One is opening a gap, $\hbar \omega_{\rm gap}=\Delta+\sqrt{\Delta^2+ (v_0 p)^2}$, and the other is gapless, $\hbar \omega_{\rm gapless}=-\Delta+\sqrt{\Delta^2+ (v_0 p)^2}$.
The angular frequency can be positive or negative (2 possibilities), and the final 2 possibilities are guided or radiative.
The forms of the wavefunctions are summarised in Table \ref{Table-III}.

\begin{table}[h]
\caption{\label{Table-III}
Possible wavefunctions in a graded-index (GRIN) fibre.
The major dependences of $\psi(r) \psi(z)$ on $r$, $z$, and $t$ are shown.
Physically unrealistic solutions are shown as n/a.
}
\begin{ruledtabular}
\begin{tabular}{lcc}
Dispersion&Guided & Radiative \\
\colrule
$\omega_0= \omega_{\rm gap} >0$ 
& 
$ {\rm e}^{i k z - i \omega_{\rm gap} t} {\rm  e}^{-\frac{r^2}{w_0^2}}$ 
& 
n/a
\\
$\omega_0= \omega_{\rm gapless} >0$ 
& 
n/a
& 
$ {\rm e}^{i k z - i \omega_{\rm gapless} t} {\rm  e}^{+\frac{r^2}{w_0^2}}$ 
\\
$\omega_0= - \omega_{\rm gap}<0$ 
& 
n/a
& 
${\rm e}^{i k z + i \omega_{\rm gapless} t} {\rm  e}^{+\frac{r^2}{w_0^2}} $ 
\\
$\omega_0= - \omega_{\rm gap} <0$ 
& 
$ {\rm e}^{i k z + i \omega_{\rm gap} t} {\rm  e}^{-\frac{r^2}{w_0^2}}$ 
& 
n/a
\end{tabular}
\end{ruledtabular}
\end{table}

If we define the time reversal operator as 
\begin{eqnarray}
\hat{\mathcal T}
t
=
-t,
\end{eqnarray}
the existence of these solutions correspond to the time reversal symmetry of the system
\begin{eqnarray}
\hat{\mathcal T}
{\rm e}^{i k z - i \omega_{\rm gap} t}
&=&
{\rm e}^{i k z + i \omega_{\rm gap} t},
\\
\hat{\mathcal T}
{\rm e}^{i k z - i \omega_{\rm gapless} t}
&=&
{\rm e}^{i k z + i \omega_{\rm gapless} t}.
\end{eqnarray}

For photonic systems without gain or loss, the time reversal symmetry is guaranteed, and we always have 2 conjugate solutions like ${\rm e}^{i k z - i \omega t}$ and ${\rm e}^{-i k z + i \omega t}$. 
These solutions are obtained by the successive applications of parity-flip and time-reversal operations
\begin{eqnarray}
\hat{\mathcal P}
\hat{\mathcal T}
{\rm e}^{i k z - i \omega t}
=
{\rm e}^{-i k z + i \omega t}.
\end{eqnarray}
The simultaneous existence of these solutions is required to guarantee the observable nature of electro-magnetic fields, since these fields must be real \cite{Yariv97,Saito20a,Saito20c}.
While the behaviours of the plane waves against these operations are rather trivial, the impacts on spin and chirality are not trivial. 
We will focus on the quantum-mechanical origin of spin for a photon in the next section.

\section{Results}
We discuss the origin of spin for a photon propagating in a GRIN fibre.
The interpretation of spin of a photon as a polarisation degree of freedom is discussed.

\subsection{$2D$ Klein-Gordon equation}
In the previous section, we have derived the $1D$ Schr\"odinger equation for a photon in a GRIN fibre along the direction of the propagation as
\begin{eqnarray}
i \hbar \partial_t \psi_z^{+}
=
-
\frac{\hbar^2}{2 m^{*}}
\Box_2 
\psi_z^{+},
\end{eqnarray}
where we have re-defined the sign of the $2D$ d'Alembertian as 
\begin{eqnarray}
\Box_2
=  
\frac{1}{v_0^2} \partial_t^2
-
\partial_z^2
\end{eqnarray}
for a comparison with a standard non-relativistic Schr\"odinger equation \cite{Baym69,Sakurai14}.
This will give an eigenvalue of $E_0^{+\pm}=\Delta \pm \sqrt{\Delta^2+ (v_0 p)^2}$.
We also obtained the complex conjugate equation for the mode propagating in the opposite direction, 
\begin{eqnarray}
- i \hbar \partial_t \psi_z^{-}
=
-
\frac{\hbar^2}{2 m^{*}}
\Box_2 
\psi_z^{-},
\end{eqnarray}
whose eigenvalue is $E_0^{-\pm}=-\Delta \pm \sqrt{\Delta^2+ (v_0 p)^2}$. 

The existence of the conjugate solution is guaranteed by the time-reversal symmetry ($t \leftrightarrow -t$) of the system

\begin{eqnarray}
\psi_z^{+}
&=&
\left(
\psi_z^{-}
\right)^{\dagger}
\\
\psi_z^{-}
&=&
\left(
\psi_z^{+}
\right)^{\dagger}.
\end{eqnarray}

Here, we are considering a coherent ray of photons, emitted from a laser source.
The mode is confined in a GRIN fibre, such that the original rotational and translational symmetries in free space are broken.
The shift of the energy of $\pm \Delta$ could be interpreted as an energy shift of the vacuum upon lasing, because the laser operation is similar to the Bose-Einstein condensation with macroscopic number of photons occupying the same state.
According to the theory of the superconductivity \cite{Bardeen57,Schrieffer71}, the energy-shift for the superconducting condensation is ${\mathcal N}_{\rm e} D(E_{\rm F})|\Delta|^2/2$, where ${\mathcal N}_{\rm e}$ is the number of electrons, $D(E_{\rm F})$ is the density of electron per energy, and $|\Delta|$ is the energy gap. 
For electrons, due to the Pauli's exclusion principle, the number of electrons affected by the superconducting paring interaction is limited to those electrons near the Fermi energy, $E_{\rm F}$.
On the other hand, photons are Bose particles, such that they can occupy the same state.
Consequently, we expect the total energy change of ${\mathcal N}|\Delta|$ for photons in a laser, where ${\mathcal N}$ is the number of photons in a cavity.
In order to consider this energy-shift by condensation, we consider a unitary transformation of the wavefunction
\begin{eqnarray}
\psi_z
=
{\rm e}^{\frac{i}{\hbar} \Delta t}
\psi_z^{+}, 
\end{eqnarray}
which yields
\begin{eqnarray}
\partial_t \psi_z^{+}
&=&
-
\frac{i}{\hbar}
\Delta
{\rm e}^{-\frac{i}{\hbar} \Delta t}
\psi_z
+
{\rm e}^{-\frac{i}{\hbar} \Delta t}
\partial_t 
\psi_z
\\
\partial_t^2 \psi_z^{+}
&=&
-
\frac{\Delta^2}{\hbar^2}
\Delta
{\rm e}^{-\frac{i}{\hbar} \Delta t}
\psi_z
-
2
\frac{i}{\hbar}
\Delta
{\rm e}^{-\frac{i}{\hbar} \Delta t}
\partial_t 
\psi_z
+
{\rm e}^{-\frac{i}{\hbar} \Delta t}
\partial_t^2 
\psi_z
\nonumber. \\
\end{eqnarray}
By inserting these into the Schr\"odinger equation, we obtain the $2D$ Klein-Goldon equation \cite{Sakurai67},
\begin{eqnarray}
\left[
\frac{1}{v_0^2} \partial_t^2
- \partial_z^2
+ \frac{m^{*2}v_0^2}{\hbar^2}
\right]
\psi_z
=
0,
\end{eqnarray}
for which the solution is
\begin{eqnarray}
\psi_z = {\rm e}^{i k z - i \omega t}
\end{eqnarray}
with the eigenvalue of 
\begin{eqnarray}
E
=
\hbar
\omega
=
\pm
\sqrt{\Delta^2+ (v_0 p)^2}.
\end{eqnarray}
The $U(1)$ unitary transformation is just a constant shift of the energy, which is considered to be attributed to the difference of vacuum energies due to the difference of symmetries between a $1D$ confined GRIN fibre and a free space.
The energy gap of $\Delta$, to be responsible for the finite effective mass of $m^{*}$ cannot be removed by this unitary transformation.

Similarly, we also consider the unitary transformation
\begin{eqnarray}
\psi_z
=
{\rm e}^{- \frac{i}{\hbar} \Delta t}
\psi_z^{-},
\end{eqnarray}
which yields
\begin{eqnarray}
\partial_t \psi_z^{-}
&=&
+
\frac{i}{\hbar}
\Delta
{\rm e}^{+\frac{i}{\hbar} \Delta t}
\psi_z
+
{\rm e}^{+\frac{i}{\hbar} \Delta t}
\partial_t 
\psi_z
\\
\partial_t^2 \psi_z^{-}
&=&
-
\frac{\Delta^2}{\hbar^2}
\Delta
{\rm e}^{+\frac{i}{\hbar} \Delta t}
\psi_z
+
2
\frac{i}{\hbar}
\Delta
{\rm e}^{+\frac{i}{\hbar} \Delta t}
\partial_t 
\psi_z
+
{\rm e}^{+\frac{i}{\hbar} \Delta t}
\partial_t^2 
\psi_z.
\nonumber
\\
\end{eqnarray}
By inserting these into the Schr\"odinger equation, we obtain the same Klein-Gordon equation,
\begin{eqnarray}
\left[
\frac{1}{v_0^2} \partial_t^2
- \partial_z^2
+ \frac{m^{*2}v_0^2}{\hbar^2}
\right]
\psi_z
=
0,
\end{eqnarray}
with the same wavefuction $\psi_z = {\rm e}^{i k z - i \omega t}$ and the energy spectrum $E=\hbar \omega=\pm \sqrt{\Delta^2+ (v_0 p)^2}$.

Alternatively, the $2D$ Klein-Gordon equation could be obtained from the spectrum of  $E=\hbar \omega=\pm \sqrt{\Delta^2+ (v_0 p)^2}$, while assuming the quantisation conditions
\begin{eqnarray}
E
&=&
\hbar \omega
=i \hbar \partial_t
\\
p
&=&
\hbar k
=
-i \hbar \partial_z
\\
\Delta
&=&
m^{*} v_0^2.
\end{eqnarray}
Inserting these into $E^2=\Delta^2+(v_0 p)^2$, we obtain
\begin{eqnarray}
- \hbar^2 \partial_t^2 \psi_z
=
\left[
m^{* 2} v_0^4
-
v_0^2 \hbar^2  \partial_z^2
\right]
\psi_z
,
\end{eqnarray}
which is the Klein-Gordon equation.
Therefore, the massive feature of the propagation for a photon in a GRIN fibre along the direction of propagation is essentially described by the Klein-Gordon equation.

\subsection{$2D$ Dirac equation}
It is well-established for an electron that the Klein-Gordon equation is not appropriate to consider the quantum-mechanical probabilistic interpretation with the conservation law \cite{Dirac30,Sakurai67}.
In order to cope with this problem, Dirac introduced spinor matrix operators, which enabled the factorisation of Klein-Gordon equation just like a primitive mathematical formula of $x^2-y^2=(x+y)(x-y)$, leading to the Dirac equation and the derivation of spin as an inherent quantum degree of freedom \cite{Dirac30,Sakurai67}.
We apply the same technique to our $2D$ Klein-Gordon equation for a photon.

We assume 
\begin{eqnarray}
&&\left(
\frac{1}{v_0} \partial_t
- \alpha_z \partial_z
- \alpha_x i\frac{m^{*} v_0 }{\hbar}
\right)
\nonumber \\
&&
\left(
\frac{1}{v_0} \partial_t
+ \alpha_z \partial_z
+ \alpha_x i\frac{m^{*} v_0 }{\hbar}
\right)
\psi_z
=
0,
\end{eqnarray}
where $\alpha_x$ and $\alpha_z$ are parameters to satisfy the Klein-Gordon equation.
After the expansion, we obtain
\begin{eqnarray}
&\left [
\frac{1}{v_0^2} \partial_t^2
-\alpha_z^2 \partial_z^2
+\frac{m^{* 2} v_0^2}{\hbar^2} \alpha_x^2
-i\frac{m^{*}v_0}{\hbar} 
( \alpha_z \alpha_x 
+
\alpha_x \alpha_z
)
\right ]
\psi_z
=
0.
\nonumber \\
\end{eqnarray}
By comparing this with the Klein-Gordon equation, we obtain
\begin{eqnarray}
&&\alpha_x^2 =1 \\
&&\alpha_z^2 =1 \\
&&\alpha_z \alpha_x 
+
\alpha_x \alpha_z
=0.
\end{eqnarray}
In order to satisfy these equations, in particular for the last anti-commutation relationship, we obtain
\begin{eqnarray}
&&\alpha_x
=
\sigma_x 
=
\left (
  \begin{array}{cc}
   0 & 1
\\
   1 & 0
  \end{array}
\right)
\\
&&\alpha_z
=
\sigma_z 
=
\left (
  \begin{array}{cc}
   1 & 0
\\
   0 & -1
  \end{array}
\right),
\end{eqnarray}
where $\sigma_i$ ($i=x,y,z$) is the Pauli spin matrices \cite{Dirac30,Baym69,Sakurai14}.
As we shall show later, the other choices of spin operators are possible, due to the rotational symmetries of spin states.

Finally, we obtain the $2D$ Dirac equation for a photon as 
\begin{eqnarray}
&&
\left(
\frac{1}{v_0} \partial_t
- \sigma_z \partial_z
- \sigma_x i\frac{m^{*} v_0 }{\hbar}
\right)
\nonumber \\
&&
\left(
\frac{1}{v_0} \partial_t
+ \sigma_z \partial_z
+ \sigma_x i\frac{m^{*} v_0 }{\hbar}
\right)
\psi_z
=
0, 
\end{eqnarray}
which could be re-written by multiplying $\hbar$ as
\begin{eqnarray}
&&
\left(
i \hbar \frac{1}{v_0} \partial_t
- i \hbar \sigma_z \partial_z
+ \sigma_x m^{*} v_0 
\right)
\nonumber \\
&&
\left(
i \hbar \frac{1}{v_0} \partial_t
+ i \hbar \sigma_z \partial_z
- \sigma_x m^{*} v_0 
\right)
\psi_z
=
0.
\end{eqnarray}
This implies 
\begin{eqnarray}
&&
\left(
i \hbar \frac{1}{v_0} \partial_t
- i \hbar \sigma_z \partial_z
+ \sigma_x m^{*} v_0 
\right)
\psi_z
=
0
\\
&&
\left(
i \hbar \frac{1}{v_0} \partial_t
+ i \hbar \sigma_z \partial_z
- \sigma_x m^{*} v_0 
\right)
\psi_z
=
0.
\end{eqnarray}

The wavefunction of the $2D$ Dirac equation is described by the 2-component spinor representation as 
\begin{eqnarray}
\psi_z
=
{\rm e}^{i k z - i \omega t}
\left (
  \begin{array}{c}
   \chi_{\uparrow} 
\\
   \chi_{\downarrow} 
  \end{array}
\right),
\end{eqnarray}
where we the spin up and down components correspond to the left and right circular polarisation states \cite{Saito20a}, and the spinor components are
\begin{eqnarray}
\chi_{\uparrow}
&=&
\langle \ \uparrow | 
\psi_{z} \rangle
\\
\chi_{\downarrow}
&=&
\langle \ \downarrow | 
\psi_{z} \rangle.
\end{eqnarray}
In this way, we have derived spin of a photon, which is assigned for the polarisation degree of freedom, from the Helmholtz equation by assuming the proper factorisation of the corresponding $2D$ Dirac equation. 

By inserting the plane wave solution along the direction of the propagation ($z$), we obtain
\begin{eqnarray}
\left(
i \hbar \partial_t
+v_0 p \sigma_z 
+ \sigma_x m^{*} v_0^2 
\right)
\psi_z
&=&
0
\\
\left(
i \hbar  \partial_t
- v_0 p \sigma_z 
- \sigma_x m^{*} v_0^2 
\right)
\psi_z
&=&
0
\end{eqnarray}
which are equations of motion to describe the time evolution of spin.
Here, we put $\xi = v_0 p $ and $\Delta = m^{*} v_0^{2} $, and they become
\begin{eqnarray}
i \hbar \partial_t
\psi_z
&=&
- 
{\bf h}
\cdot
{\boldsymbol \sigma}
\psi_z
\\
i \hbar \partial_t
\psi_z
&=& 
{\bf h}
\cdot
{\boldsymbol \sigma}
\psi_z,
\end{eqnarray}
where the effective {\it "magnetic"} field is ${\bf h}=(\Delta, 0, \xi)$ in the unit of energy, and the spin/polarisation vector is ${\boldsymbol \sigma}=(\sigma_x, \sigma_y, \sigma_z)$.
It is interesting to be aware that the 2 conjugate equations are obtained in pairs, which correspond to the opposite magnetic-field each other.
This is guaranteed by the time reversal symmetry of the system, such that we should have 2 solutions, which are conjugate each other.

In the matrix form, the equations of motion are re-written as
\begin{eqnarray}
\hbar \omega
\left (
  \begin{array}{c}
   \chi_{\uparrow} 
\\
   \chi_{\downarrow} 
  \end{array}
\right)
&=&
- 
\left (
  \begin{array}{cc}
   \xi & \Delta 
\\
   \Delta & - \xi 
  \end{array}
\right)
\left (
  \begin{array}{c}
   \chi_{\uparrow} 
\\
   \chi_{\downarrow} 
  \end{array}
\right)
\\
\hbar \omega
\left (
  \begin{array}{c}
   \chi_{\uparrow} 
\\
   \chi_{\downarrow} 
  \end{array}
\right)
&=&
+ 
\left (
  \begin{array}{cc}
   \xi & \Delta 
\\
   \Delta & - \xi 
  \end{array}
\right)
\left (
  \begin{array}{c}
   \chi_{\uparrow} 
\\
   \chi_{\downarrow} 
  \end{array}
\right),
\end{eqnarray}
whose eigenvalues are $E_{\pm}=\pm \sqrt{\xi^2 + \Delta^2}$.

We define the Hamiltonian for left and right polarising state as
\begin{eqnarray}
H_{\rm L}
&=&
{\bf h}
\cdot
{\bf \sigma}
= 
\left (
  \begin{array}{cc}
   \xi & \Delta 
\\
   \Delta & - \xi 
  \end{array}
\right)
\\
H_{\rm R}
&=&
- 
{\bf h}
\cdot
{\bf \sigma}
=
- 
\left (
  \begin{array}{cc}
   \xi & \Delta 
\\
   \Delta & - \xi 
  \end{array}
\right),
\end{eqnarray}
then, the $2D$ Dirac equation can be re-written as
\begin{eqnarray}
\left[
i \hbar \partial_t
- H_{\rm L}
\right]
\left[
i \hbar \partial_t
- H_{\rm R}
\right]
\psi_z
=
0.
\end{eqnarray}
Equivalently, we can also re-write
\begin{eqnarray}
\left[
i \hbar \partial_t
- H
\right]
\left[
- i \hbar \partial_t
- H
\right]
\psi_z
=
0,
\end{eqnarray}
where $H=H_{\rm L}$, emphasising the time-reversal symmetry and the Hermitian nature of the Hamiltonian $H^{\dagger}=H$.

\subsection{Polarisation state described by the $2D$ Dirac equation: The simplest example}
Now, it is ready to discuss the polarisation state derived by the $2D$ Dirac equation.
Here, we consider a free propagation limit of $\Delta=0$.
In this case, the Hamiltonians become
\begin{eqnarray}
H_{\rm L}
&=& 
\left (
  \begin{array}{cc}
   \xi & 0
\\
   0 & - \xi 
  \end{array}
\right) 
\\
H_{\rm R}
&= &
-
\left (
  \begin{array}{cc}
   \xi & 0
\\
   0 & - \xi 
  \end{array}
\right).
\end{eqnarray}
For the standard particle state along the time evolution of $t>0$, the solutions are given by with $\hbar \omega= \xi > 0$, and we obtain
\begin{eqnarray}
|\uparrow \ \rangle
=
|
{\rm L}
\rangle
=
\left (
  \begin{array}{c}
1\\
0
\end{array}
\right)
\end{eqnarray}
for $H_{\rm L}$ and
\begin{eqnarray}
|\downarrow \ \rangle
=
|
{\rm R}
\rangle
=
\left (
  \begin{array}{c}
0\\
1
\end{array}
\right)
\end{eqnarray}
for $H_{\rm R}$.
We can construct the arbitrary polarisation state by the superposition state of these states, $|{\rm L}\rangle$ and $|{\rm R}\rangle$.
The energy of $|{\rm L}\rangle$ and $|{\rm R}\rangle$ are the same, such that the superposition states of these states are also the same.
For example, we obtain linearly polarised states along horizontal (H), vertical (V), diagonal (D), and anti-diagonal (A) directions as
\begin{eqnarray}
|\ {\rm H} \ \rangle
&=&
\frac{1}{\sqrt{2}}
\left (
   \begin{array}{c}
   1  \\
   1 
  \end{array}
\right)
\\
|\ {\rm V} \ \rangle
&=&
\frac{-i}{\sqrt{2}}
\left (
   \begin{array}{c}
   1  \\
   -1 
  \end{array}
\right)
\\
|\ {\rm D} \ \rangle
&=&
\frac{{\rm e}^{-\frac{\pi}{4}i}}{\sqrt{2}}
\left (
   \begin{array}{c}
   1  \\
   i 
  \end{array}
\right)
\\
|\ {\rm A} \ \rangle
&=&
\frac{{\rm e}^{\frac{\pi}{4}i}}{\sqrt{2}}
\left (
   \begin{array}{c}
   1  \\
   -i 
  \end{array}
\right),
\end{eqnarray}
respectively.

We can confirm that the superposition state is actually the steady state of the $2D$ Dirac equation without the time evolution.
For example, for the horizontally polarised state, the wavefunction becomes
\begin{eqnarray}
\psi_z (t)
=
\langle t | {\rm H} \rangle
=
\frac{{\rm e}^{ikz- i \omega t}}{\sqrt{2}}
\left (
  \begin{array}{c}
   1 
\\
   1
  \end{array}
\right),
\end{eqnarray}
where $\hbar \omega = \xi $.
Then, we confirm it satisfies the $2D$ Dirac equation as
\begin{eqnarray}
&&
\left[
i \hbar \partial_t
- H_{\rm L}
\right]
\left[
i \hbar \partial_t
- H_{\rm R}
\right]
\psi_z(t)
\nonumber \\
&&
=
\left[
i \hbar \partial_t
- 
\left (
  \begin{array}{cc}
   \xi & 0
\\
   0 & - \xi 
  \end{array}
\right)
\right]
\left[
i \hbar \partial_t
+
\left (
  \begin{array}{cc}
   \xi & 0
\\
   0 & - \xi 
  \end{array}
\right)
\right]
\frac{{\rm e}^{ikz- i \omega t}}{\sqrt{2}}
\left (
  \begin{array}{c}
   1 
\\
   1
  \end{array}
\right)
\nonumber \\
&&
=
\frac{{\rm e}^{ikz- i \omega t}}{\sqrt{2}}
\xi^2
\left[
1- 
\left (
  \begin{array}{cc}
   1 & 0
\\
   0 & - 1 
  \end{array}
\right)
\right]
\left[
1
+
\left (
  \begin{array}{cc}
   1 & 0
\\
   0 & - 1 
  \end{array}
\right)
\right]
\left (
  \begin{array}{c}
   1 
\\
   1
  \end{array}
\right)
\nonumber \\
&&
=
\frac{{\rm e}^{ikz- i \omega t}}{\sqrt{2}}
\xi^2
\left[
1- 
\left (
  \begin{array}{cc}
   1 & 0
\\
   0 & - 1 
  \end{array}
\right)
\right]
\left[
\left (
  \begin{array}{c}
   1 
\\
   1
  \end{array}
\right)
+
\left (
  \begin{array}{c}
   1 
\\
   -1
  \end{array}
\right)
\right]
\nonumber \\
&&
=
\frac{{\rm e}^{ikz- i \omega t}}{\sqrt{2}}
\xi^2
\left[
\left (
  \begin{array}{c}
   2 
\\
   0
  \end{array}
\right)
-
\left (
  \begin{array}{c}
   2 
\\
   0
  \end{array}
\right)
\right]
=
0.
\end{eqnarray}
We have previously shown that the quantum-mechanical average of the spin operators correspond to the Stokes parameters \cite{Saito20a}.
For the horizontally polarised state, it becomes
\begin{eqnarray}
\langle
{\bf S}
\rangle
&=&
\left (
  \begin{array}{c}
\langle{\bf \sigma_x}\rangle
\\
\langle{\bf \sigma_y}\rangle
\\
\langle{\bf \sigma_z}\rangle
  \end{array}
\right)
\\
&=&
\left (
  \begin{array}{c}
\psi_z ^{*}(t){\bf \sigma_x} \psi_z (t)
\\
\psi_z ^{*}(t){\bf \sigma_y} \psi_z (t)
\\
\psi_z ^{*}(t){\bf \sigma_z} \psi_z (t)
  \end{array}
\right)
\\
&=&
\left (
  \begin{array}{c}
   1 
\\
   0
\\
  0
  \end{array}
\right),
\end{eqnarray}
which is independent on $t$.

Similarly, the vertically polarised state becomes
\begin{eqnarray}
\psi_z (t)
=
\langle t | {\rm V} \rangle
=
-i\frac{{\rm e}^{ikz- i \omega t}}{\sqrt{2}}
\left (
  \begin{array}{c}
   1 
\\
   -1
  \end{array}
\right),
\end{eqnarray}
which satisfy
\begin{eqnarray}
&&
\left[
i \hbar \partial_t
- H_{\rm L}
\right]
\left[
i \hbar \partial_t
- H_{\rm R}
\right]
\psi_z(t)
\nonumber \\
&&
=
\left[
i \hbar \partial_t
- 
\left (
  \begin{array}{cc}
   \xi & 0
\\
   0 & - \xi 
  \end{array}
\right)
\right]
\left[
i \hbar \partial_t
+
\left (
  \begin{array}{cc}
   \xi & 0
\\
   0 & - \xi 
  \end{array}
\right)
\right]
\frac{-i {\rm e}^{ikz- i \omega t}}{\sqrt{2}}
\left (
  \begin{array}{c}
   1 
\\
   -1
  \end{array}
\right)
\nonumber \\
&&
=
\frac{-i {\rm e}^{ikz- i \omega t}}{\sqrt{2}}
\xi^2
\left[
1- 
\left (
  \begin{array}{cc}
   1 & 0
\\
   0 & - 1 
  \end{array}
\right)
\right]
\left[
1
+
\left (
  \begin{array}{cc}
   1 & 0
\\
   0 & - 1 
  \end{array}
\right)
\right]
\left (
  \begin{array}{c}
   1 
\\
   -1
  \end{array}
\right)
\nonumber \\
&&
=
\frac{-i {\rm e}^{ikz- i \omega t}}{\sqrt{2}}
\xi^2
\left[
1- 
\left (
  \begin{array}{cc}
   1 & 0
\\
   0 & - 1 
  \end{array}
\right)
\right]
\left[
\left (
  \begin{array}{c}
   1 
\\
   -1
  \end{array}
\right)
+
\left (
  \begin{array}{c}
   1 
\\
   1
  \end{array}
\right)
\right]
\nonumber \\
&&
=
\frac{-i {\rm e}^{ikz- i \omega t}}{\sqrt{2}}
\xi^2
\left[
\left (
  \begin{array}{c}
   2 
\\
   0
  \end{array}
\right)
-
\left (
  \begin{array}{c}
   2 
\\
   0
  \end{array}
\right)
\right]
=
0
\end{eqnarray}
We also confirm that the time independent polarisation state
\begin{eqnarray}
\langle
{\bf S}
\rangle
=
\left (
  \begin{array}{c}
   -1 
\\
   0
\\
  0
  \end{array}
\right).
\end{eqnarray}

\begin{figure}[h]
\begin{center}
\includegraphics[width=7cm]{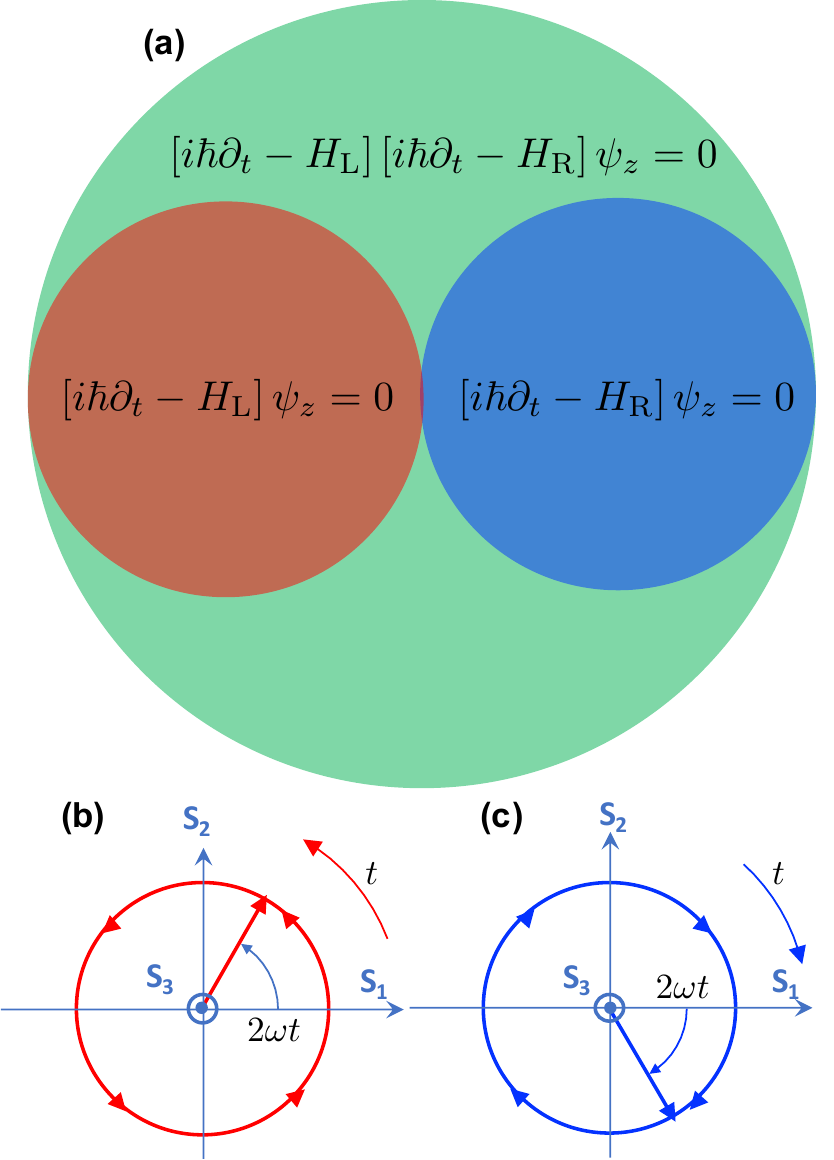}
\caption{
Dirac equation for photons and de-coupling of the equation.
(a) Schematic image of the Hilbert space spanned by the solution of the equations.
If the de-coupled equation is satisfied, the Dirac equation is automatically satisfied. 
However, there exists solutions of the Dirac equation, without satisfying the de-coupled ones.
For example, in the chiral representation, where  $ | {\rm L} \rangle =(1,0)$ and $ | {\rm R} \rangle =(0,1)$ are eigenstates of $H_{\rm L}$ and $H_{\rm R}$, respectively,  these states satisfy the decoupled equations.
However, the superposition state of  $ | {\rm H} \rangle \propto (1,1)$ and $ | {\rm V} \rangle \propto (1,-1)$ do not satisfy these de-coupled equations, but they satisfy the original Dirac equation.
Therefore, the de-coupling is not always justified.
(b) (c) Examples of incoherent states, which will not be realised.
(b) The time evolution, expected solely from $H_{\rm L}$.
If we make a superposition state of among states with positive energy $\hbar \omega>0$ and negative energy $-\hbar \omega<0$, the polarisation state becomes rapidly rotating at $2\omega$ towards the left circulation over time, seen from the top of the $S_3$ axis.
(c) The rapid rotation of polarisation state, circulating to the right solely from $H_{\rm R}$.
}
\end{center}
\end{figure}

Here, we must emphasize the decoupling of the $2D$ Dirac equation into 2 equations
\begin{eqnarray}
\left[
i \hbar \partial_t
- H_{\rm L}
\right]
\psi_z
&=&
0
\\
\left[
i \hbar \partial_t
- H_{\rm R}
\right]
\psi_z
&=&
0
\end{eqnarray}
is not always true (Fig. 2).
In fact, the polarisation states $\langle t | {\rm H} \rangle$ and  $\langle t | {\rm V} \rangle$ do not satisfy these equations at all, individually, while they satisfy the $2D$ Dirac equation (Fig. 2 (a)) in the form of the successive applications of operators as a product,  
$\left[
i \hbar \partial_t
- H_{\rm L}
\right]
\left[
i \hbar \partial_t
- H_{\rm R}
\right]$.

This could be confirmed by the example of a problem with the time evolution solely from $H_{\rm L}$.
In this case the 2 eigenvalues of $\hbar \omega=\xi$ and $-\hbar \omega=-\xi$ are assigned for $\langle t | {\rm L} \rangle$ and $\langle t | {\rm R} \rangle$, respectively. 
Then, the wavefunction at time of $t$ is given by
\begin{eqnarray}
\psi_z (t)
=
\langle t | \psi_z \rangle
=
{\rm e}^{- i \omega t}
C_{\rm L}
\langle 0 | {\rm L} \rangle
+
{\rm e}^{+ i \omega t}
C_{\rm R}
\langle 0 | {\rm R} \rangle,
\nonumber \\
\end{eqnarray}
where $C_{\rm L}$ and $C_{\rm R}$ are complex coefficients determined by the initial condition ($t=0$), which is given by
\begin{eqnarray}
\psi_z (0)
=
C_{\rm L}
\langle 0 | {\rm L} \rangle
+
C_{\rm R}
\langle 0 | {\rm R} \rangle
=
\left (
  \begin{array}{c}
C_{\rm L}\\
C_{\rm R}
\end{array}
\right).
\end{eqnarray}
Assuming the initial state is $\psi_z (0)=\langle t | {\rm H} \rangle$, we obtain
\begin{eqnarray}
\psi_z (t)
=
\frac{
  {\rm e}^{- i \omega t}
  }
  {\sqrt{2}}
\langle 0 | {\rm L} \rangle
+
\frac{
  {\rm e}^{+ i \omega t}
  }
  {\sqrt{2}}
\langle 0 | {\rm R} \rangle
=
\frac{1}{\sqrt{2}}
\left (
  \begin{array}{c}
{\rm e}^{- i \omega t}\\
{\rm e}^{+ i \omega t}
\end{array}
\right),
\nonumber \\
\end{eqnarray}
which means that the state is not an steady state.
In fact, the expectation values of the spin operators become 
\begin{eqnarray}
\langle \sigma_x \rangle
&=&
\frac{1}{2}
\left (
  \begin{array}{cc}
{\rm e}^{+ i \omega t} & 
{\rm e}^{- i \omega t}
\end{array}
\right)
\left (
  \begin{array}{cc}
0 & 1 \\
1 & 0
\end{array}
\right)
\left (
  \begin{array}{c}
{\rm e}^{- i \omega t}\\
{\rm e}^{+ i \omega t}
\end{array}
\right) 
\nonumber \\
&=&
\cos
\left( 
2 \omega t
\right)
\\
\langle \sigma_y \rangle
&=&
\frac{1}{2}
\left (
  \begin{array}{cc}
{\rm e}^{+ i \omega t} & 
{\rm e}^{- i \omega t}
\end{array}
\right)
\left (
  \begin{array}{cc}
0 & -i \\
i & 0
\end{array}
\right)
\left (
  \begin{array}{c}
{\rm e}^{- i \omega t}\\
{\rm e}^{+ i \omega t}
\end{array}
\right) 
\nonumber \\
&=&
\sin
\left( 
2 \omega t
\right)
\\
\langle \sigma_y \rangle
&=&
\frac{1}{2}
\left (
  \begin{array}{cc}
{\rm e}^{+ i \omega t} & 
{\rm e}^{- i \omega t}
\end{array}
\right)
\left (
  \begin{array}{cc}
1 & 0 \\
0 & -1
\end{array}
\right)
\left (
  \begin{array}{c}
{\rm e}^{- i \omega t}\\
{\rm e}^{+ i \omega t}
\end{array}
\right) 
\nonumber \\
&=&
0.
\end{eqnarray}
Therefore, the polarisation state become
\begin{eqnarray}
\langle
{\bf S}
\rangle
=
\left (
  \begin{array}{c}
   \cos (2 \omega t) 
\\
   \sin (2 \omega t) 
\\
  0
  \end{array}
\right),
\end{eqnarray}
which describes the incoherent left circulation (Fig. 2 (b)).
Obviously, this is in contradiction with the fact that the horizontally polarised state is maintained in a free space. 
Similarly, if we use $H_{\rm R}$ instead of $H_{\rm L}$, we found the opposite right circulation in the polarisation state (Fig. 2 (c)) as
\begin{eqnarray}
\langle
{\bf S}
\rangle
=
\left (
  \begin{array}{c}
   \cos (-2 \omega_1 t) 
\\
   \sin (-2 \omega_1 t) 
\\
  0
  \end{array}
\right).
\end{eqnarray}
The problem arose, because of the incoherent superposition between states with the positive and negative energies, which would not be realised.
The time evolution of these states will be opposite, so that we cannot consider the superposition states.
As far as we consider the states with the proper time evolution ($t>0$) and the superposition state of these state, we can de-couple the product.

\subsection{Rotation in spin state for photons: BCS-Anderson theory and Bogoliubov transformation}
In the the previous sections, we have derived the fundamental $2D$ Dirac equation and the spin of photons for describing the polarisation state for a coherent ray of photons propagating in a GRIN fibre.
In particular, the Hamiltonian we obtained has exactly the same structure of the BCS theory of superconductivity \cite{Bardeen57,Schrieffer71}.
At the early stage of the theory of superconductivity, Anderson also identified that the proposed variational state is essentially equivalent to a 2-level system, described by the superposition state between paired states and empty states, similar to the analogy to the spin density wave\cite{Anderson58}.
It is also identified by Bogoljubov\cite{Bogoljubov58} that the quasi-particle states are transferred to the different state with the energy gap upon the superconducting phase transition, and the unitary transformation is known as Bogoljubov transformation\cite{Bogoljubov58}.
Mathematically, the highlight is simply described by the diagonalisation of $2 \times 2$ matrix.
Our Hamiltonian also has the same structure, described by a 2-level system, because of the nature of $SU(2)$ spin state of a photon \cite{Saito20a}.
Therefore, we employ the BCS-Anderson theory and Bogoljubov transformation for understanding the polarisation state of photons.

\begin{figure}[h]
\begin{center}
\includegraphics[width=6cm]{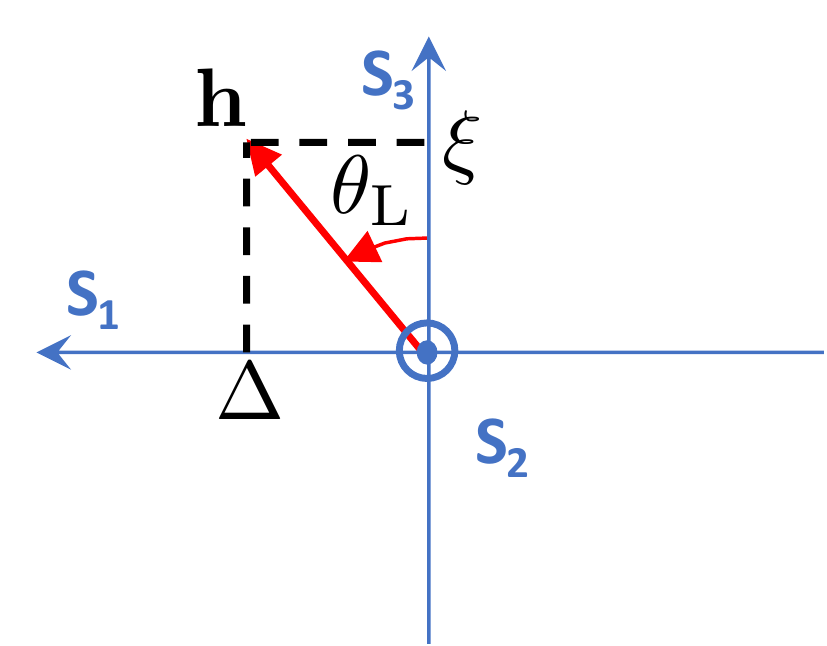}
\caption{
Poincar\'e sphere, described by Stokes parameters, ${\bf S}=(S_1,S_2,S_3)$.
The effective magnetic field ${\bf h}=(\Delta,0,\xi)$ is applied to change the polarisation state.
The original state of $| {\rm L} \langle$ at $\Delta=0$ is located at $(0,0,1)$
The eigenvalue for the proper time evolution is the state, pointing along ${\bf h}$, which is rotated along $S_2$ with the amount of $\theta_{\rm L}$.
}
\end{center}
\end{figure}

In the previous section, we used a chiral representation, for which left and right circularly polarised states are assigned to be $ | {\rm L} \rangle =(1,0)$ and $ | {\rm R} \rangle =(0,1)$ and the corresponding Stokes parameters of these states are ${\bf S}=(0,0,1)$ and ${\bf S}=(0,0,-1)$, respectively \cite{Saito20a}.
In the free space at $\Delta=0$, the polarisation eigenstate for $H_{\rm L}$ with the energy of $\hbar \omega=\xi$ was located at ${\bf S}=(0,0,1)$ and the effective magnetic field was ${\bf h}=(0,0,\xi)$, pointing along $S_3$ (Fig. 3).
We consider what happens, if the confinement energy is finite, $\Delta \neq 0$.
We use the $SU(2)$ Lie algebra to obtain the spinor wavefunction \cite{Saito20a}. 

We consider the proper time evolution by $H_{\rm L}$ with the positive energy state, which is the state, pointing along the direction of the effective magnetic field, ${\bf h}=(\Delta,0,\xi)$, in the Poincar\'e sphere (Fig. 3).
The rotation operator along $S_2$ is given by 
\begin{eqnarray}
\mathcal{D}_2 (\theta_{\rm L})
&=&
\mathcal{D}(\hat{\bf y},\theta_{\rm L})
=\exp 
\left (
  -\frac{i \sigma_y \theta_{\rm L}}{2}
\right)
\nonumber \\
&=&
{\bf 1} \cos \left( \frac{\theta_{\rm L}}{2} \right) 
-i \sigma_y \sin \left( \frac{\theta_{\rm L}}{2} \right)
\nonumber \\
&=&
\left (
  \begin{array}{cc}
    \cos \left( \frac{\theta_{\rm L}}{2} \right)  & -\sin \left( \frac{\theta_{\rm L}}{2} \right)  \\
    \sin \left( \frac{\theta_{\rm L}}{2} \right)  &  \cos \left( \frac{\theta_{\rm L}}{2} \right)
  \end{array}
\right)
\nonumber \\
&=&
\left (
  \begin{array}{cc}
    u_{\rm L}  & -v_{\rm L}  \\
    v_{\rm L}   &  u_{\rm L} 
  \end{array}
\right),
\end{eqnarray}
where $\theta_{\rm L}$ is the angle of the rotation, ${\bf 1}$ is the $2 \times 2$ unit matrix in the spinor representation, $u_{\rm L}=\cos (\theta_{\rm L}/2)$, and $u_{\rm R}=\sin (\theta_{\rm R}/2)$.
The normalisation condition,
\begin{eqnarray}
|u_{\rm L}|^2 + |v_{\rm L}|^2 =1,
\end{eqnarray}
is already satisfied.

In general for quantum mechanics, the matrix element of the Hamiltonian should not depend on the choice of the basis \cite{Baym69,Sakurai14}.
Therefore, the unitary transformation, including a rotation, should preserve the matrix element.
For example, if we consider a state given by $\langle z_i |$ and $|z_j \rangle$, the matrix element of the Hamiltonian is
\begin{eqnarray}
&&\langle z_i | H | z_j \rangle
\nonumber \\
&&=
\langle z_i | 
\mathcal{D}_y(\theta_{\rm L})
\mathcal{D}_y^{\dagger}(\theta_{\rm L})
H 
\mathcal{D}_y(\theta_{\rm L})
\mathcal{D}_y^{\dagger}(\theta_{\rm L})
| z_j \rangle \\
&&=
\langle z{'}_i | 
\mathcal{D}_y^{\dagger}(\theta_{\rm L})
H 
\mathcal{D}_y(\theta_{\rm L})
| z{'}_j \rangle \\
&&=
\langle z{'}_i | 
H '
| z{'}_j \rangle. 
\end{eqnarray}
Therefore, the unitary transformation will rotate the basis state in the opposite way,
\begin{eqnarray}
| z{'} \rangle 
=
\mathcal{D}^{\dagger}_y(\theta_{\rm L})
| z \rangle , 
\end{eqnarray}
while the Hamiltonian is transferred as $\mathcal{D}_y^{\dagger}(\theta_{\rm L})
H_{\rm L} 
\mathcal{D}_y(\theta_{\rm L})
$.
For our case, this becomes
\begin{eqnarray}
&&
\mathcal{D}_y^{\dagger}(\theta_{\rm L})
H_{\rm L} 
\mathcal{D}_y(\theta_{\rm L})
\nonumber \\
&&
=
\left (
  \begin{array}{cc}
     \cos \left( \frac{\theta_{\rm L}}{2} \right)  &  \sin \left( \frac{\theta_{\rm L}}{2} \right)  \\
    -\sin \left( \frac{\theta_{\rm L}}{2} \right)  &  \cos \left( \frac{\theta_{\rm L}}{2} \right)
  \end{array}
\right)
\left (
  \begin{array}{cc}
   \xi & \Delta 
\\
   \Delta & - \xi 
  \end{array}
\right)
\left (
  \begin{array}{cc}
    \cos \left( \frac{\theta_{\rm L}}{2} \right)  & -\sin \left( \frac{\theta_{\rm L}}{2} \right)  \\
    \sin \left( \frac{\theta_{\rm L}}{2} \right)  &  \cos \left( \frac{\theta_{\rm L}}{2} \right)
  \end{array}
\right)
\nonumber \\
&&
=
\left (
  \begin{array}{cc}
      \xi    \cos \theta_{\rm L} +
      \Delta \sin \theta_{\rm L} 
  &  
      -\xi    \sin \theta_{\rm L} +
       \Delta \cos \theta_{\rm L} 
\\
      -\xi    \sin \theta_{\rm L} +
       \Delta \cos \theta_{\rm L} 
  &  
      -\xi    \cos \theta_{\rm L} 
      -\Delta \sin \theta_{\rm L} 
  \end{array}
\right).
\end{eqnarray}
By eliminating the off-diagonal component, we obtain the gap equation, 
\begin{eqnarray}
\tan \theta_{\rm L}
=
\frac{\Delta}{\xi},
\end{eqnarray}
to diagonalise the Hamiltonian.
The graphical representation of $\theta_{\rm L}$ is shown in Fig. 3.
The gap equation gives
\begin{eqnarray}
\cos \theta_{\rm L}
&=&
\frac{\xi}{\sqrt{\xi^2+\Delta^2}}
\\
\sin \theta_{\rm L}
&=&
\frac{\Delta}{\sqrt{\xi^2+\Delta^2}},
\end{eqnarray}
and consequently, 
\begin{eqnarray}
\xi    \cos \theta_{\rm L} +
\Delta \sin \theta_{\rm L} 
=
\sqrt{\xi^2+\Delta^2}.
\end{eqnarray}
Thus, we diagonalised the Hamiltonian as 
\begin{eqnarray}
\mathcal{D}_y^{\dagger}(\theta_{\rm L})
H_{\rm L} 
\mathcal{D}_y(\theta_{\rm L})
&=&
\sqrt{\xi^2+\Delta^2}
\left (
  \begin{array}{cc}
1  &  
0\\
0  &  
-1  \end{array}
\right)
\\
&=&
\sqrt{\xi^2+\Delta^2}
\sigma_z.
\end{eqnarray}
Alternatively, the $2\times 2$ Hamiltonian could be simply diagonalised from the vanishing determinant as
\begin{eqnarray}
\left |
  \begin{array}{cc}
\xi-E  &  \Delta\\
\Delta  &  -\xi - E  \end{array}
\right |
=E^2 - \xi^2 -\Delta^2=0,
\end{eqnarray}
which gives the famous quasi-particle dispersion of
\begin{eqnarray}
E
=
\pm
\sqrt{
  \xi^2 + \Delta^2
  }.
\end{eqnarray}

It is also useful to note that we obtained the rotational parameters as 
\begin{eqnarray}
u_{\rm L}^2
&=&
\cos^2 \left ( \frac{\theta_{\rm L}}{2} \right)
=
\frac{1}{2}
\left (
1+
\cos \theta_{\rm L}
\right )
\\
&=&
\frac{1}{2}
\left (
1+
\frac{\xi}{\sqrt{\xi^2+\Delta^2}}
\right )
\\
v_{\rm L}^2
&=&
\sin^2 \left ( \frac{\theta_{\rm L}}{2} \right)
=
\frac{1}{2}
\left (
1-
\cos \theta_{\rm L}
\right )
\\
&=&
\frac{1}{2}
\left (
1-
\frac{\xi}{\sqrt{\xi^2+\Delta^2}}
\right ).
\end{eqnarray}

The spinor wavefunction for $E=\sqrt{  \xi^2 + \Delta^2  }$ is given by 
\begin{eqnarray}
u_{\uparrow}
=
\mathcal{D}_y (\theta_{\rm L})
\left (
  \begin{array}{c}
1  \\
0  
\end{array}
\right)
=
\left (
  \begin{array}{c}
u_{\rm L}  \\
v_{\rm L}  
\end{array}
\right)
=
\left (
  \begin{array}{cc}
    \cos \left( \frac{\theta_{\rm L}}{2} \right)    \\
    \sin \left( \frac{\theta_{\rm L}}{2} \right)  
  \end{array}
\right) , 
\end{eqnarray}
for the proper time evolution, while the wavefunction for the time-reversal evolution is given by
\begin{eqnarray}
u_{\downarrow}
=
\mathcal{D}_y (\theta_{\rm L})
\left (
  \begin{array}{c}
0  \\
1  
\end{array}
\right)
=
\left (
  \begin{array}{c}
-v_{\rm L}  \\
u_{\rm L}  
\end{array}
\right)
=
\left (
  \begin{array}{cc}
    - \sin \left( \frac{\theta_{\rm L}}{2} \right)  \\
    \cos \left( \frac{\theta_{\rm L}}{2} \right)    
  \end{array}
\right) ,
\nonumber \\
\end{eqnarray}
for $E= - \sqrt{  \xi^2 + \Delta^2  }$.

\begin{table}[h]
\caption{\label{Table-IV}
Summary of the solution obtained from the decoupled Dirac equation for photons with time evolution by $H_{\rm L}$.
The standard time evolution is described by the eigenstate of the positive energy, while the conjugate solution with the negative energy is obtained for the time-reversal symmetry.
The original spinor wavefunction is based on the basis with the left-circular polarised state of $|{\rm L} \rangle$ with ${\bf S}=(0,0,1)$ for $\Delta=0$, and it was rotated in the Poincar\'e sphere for $\Delta \neq 0$.
The diagonal basis is the basis to re-set the axis along the direction of the effective magnetic field.
}
\begin{ruledtabular}
\begin{tabular}{ccc}
Energy&Diagonal basis & Original basis \\
\colrule
$\sqrt{  \xi^2 + \Delta^2  }$ 
& 
$ 
|{\rm L} \rangle
=
\left (
  \begin{array}{cc}
    1  \\
    0   
  \end{array}
\right)
$ 
& 
$ 
\left (
  \begin{array}{cc}
    \cos \left( \frac{\theta_{\rm L}}{2} \right)  \\
    \sin \left( \frac{\theta_{\rm L}}{2} \right)    
  \end{array}
\right)
$ 
\\
$- \sqrt{  \xi^2 + \Delta^2  }$ 
& 
$ 
\left (
  \begin{array}{cc}
    0  \\
    1    
  \end{array}
\right)
$ 
& 
$ 
\left (
  \begin{array}{cc}
    - \sin \left( \frac{\theta_{\rm L}}{2} \right)  \\
    \cos \left( \frac{\theta_{\rm L}}{2} \right)    
  \end{array}
\right)
$ 
\end{tabular}
\end{ruledtabular}
\end{table}

The diagonalised solution is summarised in Table \ref{Table-IV}.
We have assumed that the original basis state before the rotation ($\Delta=0$) is assumed to be $|{\rm L} \rangle$ at ${\bf S}=(0,0,1)$.
Using this original basis, we obtained the spinor wavefunction by the rotation in $SU(2)$ Hilbert space.
We can also re-define this rotated state as $|{\rm L} \rangle=(1,0)$ for a new diagonal basis (Table \ref{Table-IV}) after the Bogoljubov transformation \cite{Bogoljubov58}, described above for photons.
We also obtained the conjugate solution for the time-reversal evolution.

In the strong confinement limit of $\xi \rightarrow 0$, the propagation along $z$ is limited by the heavy effective mass of $m^{*}$ and the dynamics of photons is dominated by the circular motion with orbital angular momentum and radial oscillation.
Then, the energy is dominated by $\Delta$, and the original $|{\rm L} \rangle$ would be rotated with the rotational angle of $\pi/2$ to be $|{\rm H} \rangle$ at ${\bf S}=(1,0,0)$ (Fig. 3) by $H_{\rm L}$.

Similarly, we consider the time evolution by $H_{\rm R}=-H_{\rm L}$.
The solution could be easily obtained by recognising the mapping of
\begin{eqnarray}
\Delta &\rightarrow& - \Delta
\\
 \xi &\rightarrow& -\xi,
\end{eqnarray}
from $H_{\rm L}$ to $H_{\rm R}$.
This will give the gap equation for $\theta_{\rm R}$ as
\begin{eqnarray}
\tan \theta_{\rm R}
=
\frac{\Delta}{\xi},
\end{eqnarray}
which provides
\begin{eqnarray}
\cos \theta_{\rm R}
&=&
-\frac{\xi}{\sqrt{\xi^2+\Delta^2}}
=
-
\cos \theta_{\rm L}
\\
\sin \theta_{\rm R}
&=&
-
\frac{\Delta}{\sqrt{\xi^2+\Delta^2}}
=
-
\sin \theta_{\rm L}.
\end{eqnarray}
Therefore, we obtained the relationship of 
\begin{eqnarray}
\theta_{\rm R}
=
\theta_{\rm L}
+
\pi .
\end{eqnarray}

\begin{table}[h]
\caption{\label{Table-V}
Summary of the solution obtained from the decoupled Dirac equation for photons with time evolution by $H_{\rm R}$.
There is a relationship between left and right rotation angles as $\theta_{\rm R}=\theta_{\rm L}+\pi $.
}
\begin{ruledtabular}
\begin{tabular}{ccc}
Energy&Diagonal basis & Original basis \\
\colrule
$\sqrt{  \xi^2 + \Delta^2  }$ 
& 
$ 
|{\rm R} \rangle
=
\left (
  \begin{array}{cc}
    0  \\
    1   
  \end{array}
\right)
$ 
& 
$ 
\left (
  \begin{array}{cc}
    \cos \left( \frac{\theta_{\rm R}}{2} \right)  \\
    \sin \left( \frac{\theta_{\rm R}}{2} \right)    
  \end{array}
\right)
=
\left (
  \begin{array}{cc}
    -\sin \left( \frac{\theta_{\rm L}}{2} \right)  \\
     \cos \left( \frac{\theta_{\rm L}}{2} \right)    
  \end{array}
\right)
$ 
\\
$- \sqrt{  \xi^2 + \Delta^2  }$ 
& 
$ 
-
\left (
  \begin{array}{cc}
    1  \\
    0    
  \end{array}
\right)
$ 
& 
$ 
\left (
  \begin{array}{cc}
    - \sin \left( \frac{\theta_{\rm R}}{2} \right)  \\
    \cos \left( \frac{\theta_{\rm R}}{2} \right)    
  \end{array}
\right)
=
-
\left (
  \begin{array}{cc}
    \cos \left( \frac{\theta_{\rm L}}{2} \right)  \\
    \sin \left( \frac{\theta_{\rm L}}{2} \right)    
  \end{array}
\right)
$ 
\end{tabular}
\end{ruledtabular}
\end{table}

The summary of the solutions for $H_{\rm R}$ is shown in Table \ref{Table-V}.
We obtained $|{\rm R} \rangle=(0,1)$ for a standard time evolution.
As a result, we obtained both $|{\rm L} \rangle$ and $|{\rm R} \rangle$ for the same energy of $E=\sqrt{  \xi^2 + \Delta^2  }>0$.
An arbitrary polarised state can be constructed by the superposition state of these orthogonal states \cite{Saito20a}.
Therefore, the Dirac equation can describe the time evolution of arbitrary polarised states for photons propagating in a GRIN fibre.

\subsection{Freedom to assign polarisation state}
So far, we have used a chiral representation for the basis states of $|{\rm L} \rangle$ and $|{\rm R} \rangle$ for describing the polarisation states of $(1,0)$ and $(0,1)$ \cite{Saito20a}.
We also call this basis as LR-basis\cite{Saito20a}.
The Stokes parameters are given by 
\begin{eqnarray}
{\bf S}
=
S_0
\left (
  \begin{array}{c}
    1 \\
    \langle \sigma_x \rangle \\
    \langle \sigma_y \rangle\\
    \langle \sigma_z \rangle 
  \end{array}
\right),
\end{eqnarray}
where $S_0=\hbar {\mathcal N}$ is the time-averaged total magnitude of the spin angular momentum for the number of photons of ${\mathcal N}$ in the LR-basis.
However, the original Helmholtz equation for the wavefunction of photons do not have any particular preferential direction for the polarisation states.
Consequently, we should preserve the rotational symmetry of the polarisation states. 
Therefore, the polarisation state should not be dependent on any particular choice of the basis states.

In Jones vector representation \cite{Jones41,Poincare92,Yariv97,Goldstein11,Gil16}, we use $|{\rm H} \rangle=(1,0)$ and $|{\rm V} \rangle=(0,1)$ for the basis states, and we are calling as HV-basis \cite{Saito20a}.
In this representation, the Stokes parameters are 
\begin{eqnarray}
{\bf S}
=
S_0
\left (
  \begin{array}{c}
    1 \\
    \langle \sigma_z \rangle \\
    \langle \sigma_x \rangle\\
    \langle \sigma_y \rangle 
  \end{array}
\right).
\end{eqnarray}

In our formulation, there is no difference at all, whether we should use LR or HV basis.
We can also use the diagonal basis states, composed of $|{\rm D} \rangle=(1,0)$ and $|{\rm A} \rangle=(0,1)$.
In the end, we obtain the degenerate energy states, which are mutually orthogonal. 
Arbitrary polarised states can be constructed by the superposition of these orthogonal basis states, such that the final state is not polarisation dependent.

\subsection{Pancharatnam-Berry's phase}
Here, we briefly mention about the Pancharatnam-Berry's phase \cite{Pancharatnam56,Berry84}, hidden in our derivation.
The Pancharatnam-Berry's phase is very important to consider the quantum-mechanical commutation relationship of spin operators for photons \cite{Saito20a,Saito20b,Saito20c}, and its impact on the interference \cite{Tomita86,Simon93,Bliokh09}.

In the gap equation, we only obtained $\tan \theta_{\rm L}$ or $\tan \theta_{\rm R}$ for $H_{\rm L}$ and $H_{\rm R}$, respectively.
Therefore, in addition to the solutions of $\theta_{\rm L}$ and $\theta_{\rm R}$, we also have alternative solutions of $\theta_{\rm L}+\pi$ and $\theta_{\rm R}+\pi$.
These solutions correspond to the change of $\xi \rightarrow -\xi$ and $\Delta \rightarrow - \Delta$, and therefore, they merely change the overall sign of the spinor wavefunction, $u_{\sigma} \rightarrow - u_{\sigma}$ for both polarisations of $\sigma = {\rm \uparrow}$ and $\sigma = {\rm \downarrow}$.
In the diagonal basis, these just correspond to the changes of $|{\rm L} \rangle \rightarrow -|{\rm L} \rangle$ and $|{\rm R} \rangle \rightarrow -|{\rm R} \rangle$.
This is nothing but a Pancharatnam-Berry's phase \cite{Pancharatnam56,Berry84,Saito20a}.

For the polarisation state, described by the Stokes parameters of ${\bf S}=(S_1,S_2,S_3)$, there is no difference at all by the difference of Pancharatnam-Berry's phase.
Nevertheless, this phase is observable by interference experiments \cite{Tomita86,Simon93,Bliokh09,Saito20a}.
In the BCS theory of superconductivity, this degree of freedom is hidden in a superconducting order parameter as a phase degree of freedom \cite{Nambu59,Anderson58,Goldstone62,Higgs64,Schrieffer71}.
Our formulation of the Dirac equation for photons has the same mathematical structure, and we will show the importance of the phase degree of freedom in the order parameter of $\Delta$, subsequently.

\subsection{Poincar\'e sphere}

Here, we discuss the importance of the phase in the energy gap, and its impact on the polarisation states.
We discuss the principle of rotational symmetry in polarisation states and the role of the phase of the energy gap, which corresponds to the azimuthal angle.
First, we go back to the Klein-Gordon equation and discuss the freedom to assign Pauli spin matrices.
We consider another choice of 
\begin{eqnarray}
&&
\alpha_x
=
\sigma_y 
=
\left (
  \begin{array}{cc}
   0 & -i
\\
   i & 0
  \end{array}
\right)
\\
&&
\alpha_z
=
\sigma_z 
=
\left (
  \begin{array}{cc}
   1 & 0
\\
   0 & -1
  \end{array}
\right).
\end{eqnarray}
In this choice of the gauge, we obtain the Dirac equation as 
\begin{eqnarray}
\left(
i \hbar \partial_t
- \xi \sigma_z 
- \Delta \sigma_y 
\right)
\left(
i \hbar \partial_t
+ \xi \sigma_z 
+ \Delta \sigma_y 
\right)
\psi_z
=
0,
\nonumber \\
\end{eqnarray}
using the same parameters of $\xi=v_0 p$ and $\Delta=m^{*} v_0^2$.
The only difference from the previous choice is the direction of the effective magnetic field as 
${\bf h}=( 0, \Delta, \xi)$.

\begin{figure}[h]
\begin{center}
\includegraphics[width=6cm]{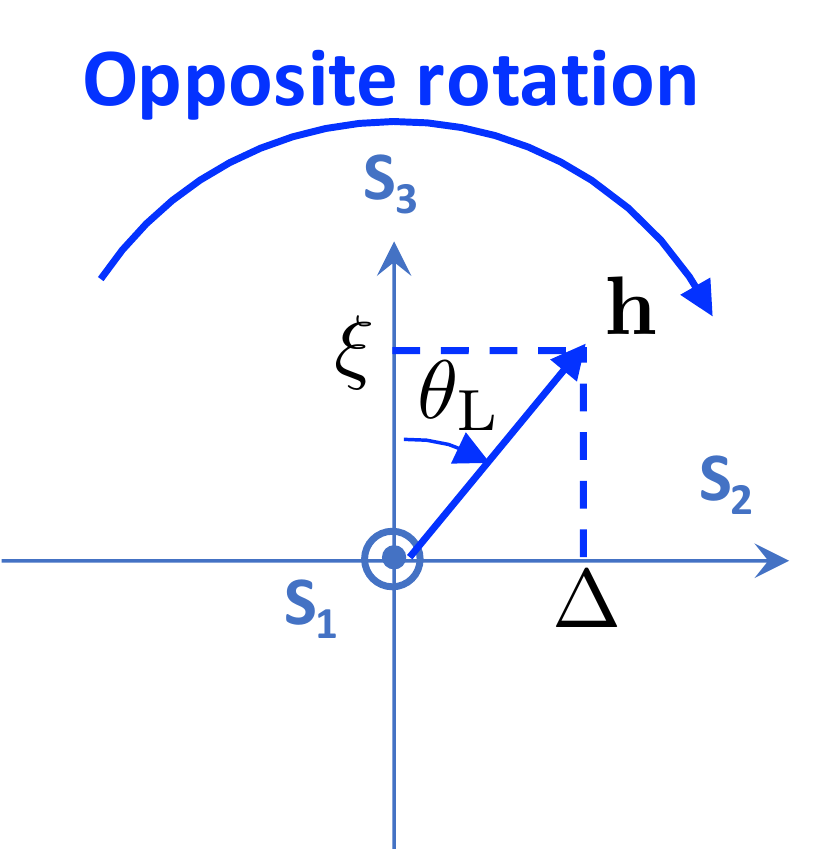}
\caption{
Another choice of the gauge for the polarisation rotation in the Poincar\'e sphere.
The effective magnetic field ${\bf h}=(0,\Delta,\xi)$ is applied to change the polarisation state.
The rotation is along $S_1$ with the amount of $- \theta_{\rm L}$, which is the opposite direction with the previous choice.
}
\end{center}
\end{figure}

In this case, for the solution of the de-coupled Dirac equation of $H_{\rm L}$, the eigenstate pointing ${\bf h}$ is obtained by the rotation along $S_1$ in the clock-wise direction, seen from the top of the $S_1$ axis with $\theta_{\rm L}$, which is given by the $SU(2)$ rotation operator, 
\begin{eqnarray}
\mathcal{D}_1 (- \theta_{\rm L})
&=&
\mathcal{D}(\hat{\bf x},-\theta_{\rm L})
=
\exp 
\left (
  +\frac{i \sigma_x \theta_{\rm L}}{2}
\right)
\nonumber \\
&=& {\bf 1} \cos \left( \frac{\theta_{\rm L}}{2} \right) 
+i \sigma_x \sin \left( \frac{\theta_{\rm L}}{2} \right)
\nonumber \\
&=&
\left (
  \begin{array}{cc}
    \cos \left( \frac{\theta_{\rm L}}{2} \right)    & i \sin \left( \frac{\theta_{\rm L}}{2} \right)  \\
    i \sin \left( \frac{\theta_{\rm L}}{2} \right)  &   \cos \left( \frac{\theta_{\rm L}}{2} \right)
  \end{array}
\right)
\nonumber \\
&=&
\left (
  \begin{array}{cc}
    u_{\rm L}  & iv_{\rm L}  \\
    iv_{\rm L}   &  u_{\rm L} 
  \end{array}
\right).
\end{eqnarray}
The rotated Hamiltonian becomes
\begin{eqnarray}
&&
\mathcal{D}_x^{\dagger}(-\theta_{\rm L})
H_{\rm L} 
\mathcal{D}_x(-\theta_{\rm L})
\nonumber \\
&=&
\left (
  \begin{array}{cc}
     \cos \left( \frac{\theta_{\rm L}}{2} \right)     &  - i \sin \left( \frac{\theta_{\rm L}}{2} \right)  \\
     -i \sin \left( \frac{\theta_{\rm L}}{2} \right)  &  \cos \left( \frac{\theta_{\rm L}}{2} \right)
  \end{array}
\right)
\left (
  \begin{array}{cc}
   \xi & - i \Delta 
\\
   i \Delta & - \xi 
  \end{array}
\right)
\nonumber \\
&&
\left (
  \begin{array}{cc}
    \cos \left( \frac{\theta_{\rm L}}{2} \right)    & i \sin \left( \frac{\theta_{\rm L}}{2} \right)  \\
    i \sin \left( \frac{\theta_{\rm L}}{2} \right)  &  \cos \left( \frac{\theta_{\rm L}}{2} \right)
  \end{array}
\right)
\nonumber \\
&=&
\left (
  \begin{array}{cc}
      \xi    \cos \theta_{\rm L} +
      \Delta \sin \theta_{\rm L} 
  &  
      i\xi    \sin \theta_{\rm L} 
      -i\Delta \cos \theta_{\rm L} 
\\
      -i\xi    \sin \theta_{\rm L} +
       i\Delta \cos \theta_{\rm L} 
  &  
      -\xi    \cos \theta_{\rm L} 
      -\Delta \sin \theta_{\rm L} 
  \end{array}
\right).
\nonumber \\
\end{eqnarray}

By eliminating the off-diagonal component, we obtain the same gap equation of 
\begin{eqnarray}
\tan \theta_{\rm L}
=
\frac{\Delta}{\xi}
\end{eqnarray}
as in the previous choice, such that the angle $\theta_{\rm L}$ is not changed.
Consequently, we obtain the same eigenvalues of $E=\pm\sqrt{  \xi^2 + \Delta^2  }$.
However, we obtain slightly different eigenfunctions
\begin{eqnarray}
u_{\uparrow}
&=&
\mathcal{D}_x(-\theta_{\rm L})
\left (
  \begin{array}{c}
1  \\
0  
\end{array}
\right)
=
\left (
  \begin{array}{c}
u_{\rm L}  \\
i v_{\rm L}  
\end{array}
\right)
\nonumber \\
&=&
\left (
  \begin{array}{cc}
    \cos \left( \frac{\theta_{\rm L}}{2} \right)    \\
    i \sin \left( \frac{\theta_{\rm L}}{2} \right)  
  \end{array}
\right)
\\
u_{\downarrow}
&=&
\mathcal{D}_x (-\theta_{\rm L})
\left (
  \begin{array}{c}
0  \\
1  
\end{array}
\right)
=
\left (
  \begin{array}{c}
i v_{\rm L}  \\
u_{\rm L}  
\end{array}
\right)
\nonumber \\
&=&
\left (
  \begin{array}{cc}
    i\sin \left( \frac{\theta_{\rm L}}{2} \right)  \\
    \cos \left( \frac{\theta_{\rm L}}{2} \right)    
  \end{array}
\right).
\end{eqnarray}
These states are orthogonal each other, as confirmed by the inner product
\begin{eqnarray}
u_{\uparrow}^{*}
u_{\downarrow}
&=&
\left (
  \begin{array}{cc}
    \cos \left( \frac{\theta_{\rm L}}{2} \right)    &
    -i \sin \left( \frac{\theta_{\rm L}}{2} \right)  
  \end{array}
\right)
\left (
  \begin{array}{cc}
    i\sin \left( \frac{\theta_{\rm L}}{2} \right)  \\
    \cos \left( \frac{\theta_{\rm L}}{2} \right)    
  \end{array}
\right)
\nonumber \\
&=&0.
\end{eqnarray}
For the standard time evolution of the energy of $E= \sqrt{  \xi^2 + \Delta^2}$, we obtained 
\begin{eqnarray}
|{\rm L} \rangle
=
\left (
  \begin{array}{cc}
    \cos \left( \frac{\theta_{\rm L}}{2} \right)    \\
    i \sin \left( \frac{\theta_{\rm L}}{2} \right)  
  \end{array}
\right)
\end{eqnarray}
in the original chiral basis, if the magnetic field is applied to ${\bf h}=(0,\Delta,\xi)$.
This is in contrast to the previous solution of 
\begin{eqnarray}
|{\rm L} \rangle
=
\left (
  \begin{array}{cc}
    \cos \left( \frac{\theta_{\rm L}}{2} \right)    \\
    \sin \left( \frac{\theta_{\rm L}}{2} \right)  
  \end{array}
\right)
\end{eqnarray}
for the field of ${\bf h}=(\Delta,0,\xi)$.

The difference of the phase of $i$, could be understood by the rotation in the $S_1-S_2$ plane, which is enabled by a rotator operator \cite{Saito20a},
\begin{eqnarray}
\mathcal{D}_3 
\left (
\frac{\pi}{2}
\right )
&=&
\exp 
\left (
  +\frac{i \sigma_z \pi}{4}
\right)
=
\left (
  \begin{array}{cc}
    {\rm e}^{-i\frac{\pi}{4}}    & 0  \\
    0  &  {\rm e}^{+i\frac{\pi}{4}} 
\end{array}
\right)
\nonumber
\\
&=&
{\rm e}^{-i\frac{\pi}{4}}
\left (
  \begin{array}{cc}
    1    & 0  \\
    0  &  i
\end{array}
\right)
=
{\rm e}^{+i\frac{\pi}{4}}
\left (
  \begin{array}{cc}
    -i    & 0  \\
    0  &  1
\end{array}
\right).
\end{eqnarray}
In fact, by applying this rotator, we confirm
\begin{eqnarray}
\mathcal{D}_3 
\left (
\frac{\pi}{2}
\right )
\left (
  \begin{array}{cc}
    \cos \left( \frac{\theta_{\rm L}}{2} \right)    \\
    \sin \left( \frac{\theta_{\rm L}}{2} \right)  
  \end{array}
\right)
=
{\rm e}^{-\frac{\pi}{4}i}
\left (
  \begin{array}{cc}
    \cos \left( \frac{\theta_{\rm L}}{2} \right)    \\
    i\sin \left( \frac{\theta_{\rm L}}{2} \right)  
  \end{array}
\right).
\end{eqnarray}
Therefore, the choice of the Pauli spin matrices are just coming from the choice of the polarisation axis in the $S_1-S_2$ plane.

In order to confirm this view, we consider a more elaborate choice of the Pauli spin matrices for the derivation of the Dirac equation as
\begin{eqnarray}
\alpha_z
&=&
\sigma_z 
=
\left (
  \begin{array}{cc}
   1 & 0
\\
   0 & -1
  \end{array}
\right)
\\
\alpha_x
&=&
\sigma_x \cos \phi + \sigma_y \sin \phi 
=
\left (
  \begin{array}{cc}
   0 & {\rm e}^{-i \phi}
\\
   {\rm e}^{i \phi} & 0
  \end{array}
\right),
\end{eqnarray}
where $\phi$ is the phase, describing the azimuthal angle in the Poincar\'e sphere (Fig. 5).
In addition to the trivial identity, $\sigma_z ^2={\bf 1}$, these choices in fact satisfy the splitting conditions, 
\begin{eqnarray}
\alpha_x^2 
&=&
\sigma_x \cos^2 \phi
+
\sigma_y \sin^2 \phi
\cos \phi \sin \phi
\left (
\sigma_x \sigma_y + \sigma_y \sigma_x
\right )
\nonumber
\\
&=&
{\bf 1}
\\
\alpha_x \sigma_z
&=&
\sigma_x \sigma_z \cos \phi
\sigma_y \sigma_z \sin \phi
=
-\sigma_z
\left (
\sigma_x \cos \phi \sigma_y \sin \phi
\right )
\nonumber
\\
&=&
-\sigma_z
\alpha_x.
\end{eqnarray}

In this gauge, the Dirac equation becomes
\begin{eqnarray}
& &
\left [ 
i \hbar \partial_t
- \xi \sigma_z 
-  \Delta \left ( \cos \phi \sigma_x + \sin \phi \sigma_y \right)
\right] 
\nonumber \\
&&
\left [ 
i \hbar \partial_t
+ \xi \sigma_z 
+  \Delta \left ( \cos \phi \sigma_x + \sin \phi \sigma_y \right)
\right] 
\psi_z
=
0.
\end{eqnarray}
Thus, the magnetic field is ${\bf h}=( \Delta \cos \phi, \Delta \sin \phi, \xi)$.
The Hamiltonian for the left state is
\begin{eqnarray}
H_{\rm L}
=
{\bf h}
\cdot
{\bf \sigma}
= 
\left (
  \begin{array}{cc}
   \xi & \Delta {\rm e}^{- i \phi}
\\
   \Delta {\rm e}^{i \phi} & - \xi 
  \end{array}
\right),
\end{eqnarray}
which is in fact Hermite
\begin{eqnarray}
H_{\rm L}^{\dagger}
=
H_{\rm L}
=H.
\end{eqnarray}
Similarly, the Hamiltonian for the right state is
\begin{eqnarray}
H_{\rm R}
=
-H_{\rm L}
=
-{\bf h}
\cdot
{\bf \sigma}
= 
-\left (
  \begin{array}{cc}
   \xi & \Delta {\rm e}^{- i \phi}
\\
   \Delta {\rm e}^{i \phi} & - \xi 
  \end{array}
\right).
\end{eqnarray}
We can now identify that the introduced angle of $\phi$ is actually the phase of the order parameter, $\Delta {\rm e}^{i \phi}$.

\begin{figure}[h]
\begin{center}
\includegraphics[width=7cm]{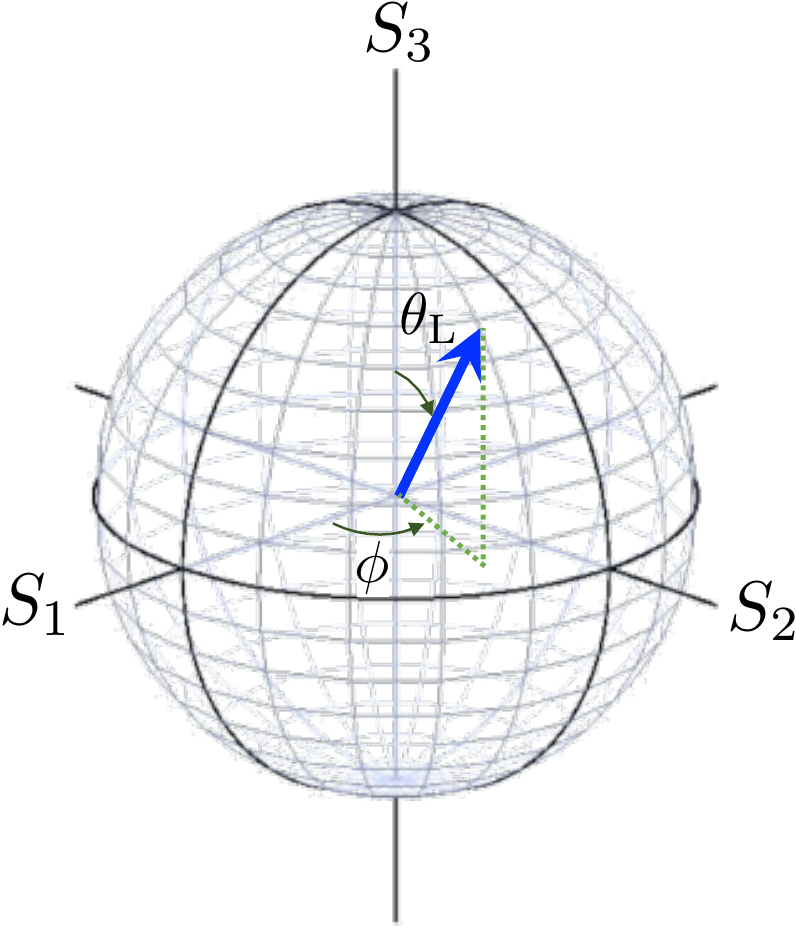}
\caption{
Stokes parameters in Poincar\'e sphere.
The effective magnetic field, ${\bf h}=( \Delta \cos \phi, \Delta \sin \phi, \xi)$, is pointing to the direction of the polarisation state.
The phase ($\phi$) of the order parameter, $\Delta {\rm e}^{i \phi} \neq 0$, is the azimuthal angle of the polarisation state.
The polar angle ($\theta_{\rm L}$) is determined by the gap equation of $\tan (\theta_{\rm L})=\Delta/ \xi$, which is determined by the ratio of the energy for the confinement energy (orbital and radial contributions) and the kinetic energy along the direction of the propagation.
}
\end{center}
\end{figure}

In order to eliminate the impact of the phase, we can rotate the polarisation state in the Poincar\'e sphere (Fig. 5).
To obtain the left circularly polarised state for $\Delta {\rm e}^{i \phi} \neq 0$ for $H_{\rm L}$, first we rotate along the $S_3$ axis by
\begin{eqnarray}
\mathcal{D}_3 
\left (
\phi
\right )
&=
\exp 
\left (
  -\frac{i \sigma_z \phi}{2}
\right) \\
&=
\left (
  \begin{array}{cc}
    {\rm e}^{-i\frac{\phi}{2}}    & 0  \\
    0  &  {\rm e}^{+i\frac{\phi}{2}} 
\end{array}
\right).
\end{eqnarray}
The rotated Hamiltonian would be 
\begin{eqnarray}
&&\mathcal{D}_z^{\dagger}(\phi) H_{\rm L}  \mathcal{D}_z(\phi)
\nonumber \\
&&=
\left (
  \begin{array}{cc}
    {\rm e}^{+i\frac{\phi}{2}}    & 0  \\
    0  &  {\rm e}^{-i\frac{\phi}{2}} 
\end{array}
\right)
\left (
  \begin{array}{cc}
   \xi & \Delta {\rm e}^{- i \phi}
\\
   \Delta {\rm e}^{i \phi} & - \xi 
  \end{array}
\right)
\left (
  \begin{array}{cc}
    {\rm e}^{-i\frac{\phi}{2}}    & 0  \\
    0  &  {\rm e}^{+i\frac{\phi}{2}} 
\end{array}
\right)
\nonumber \\
&&= 
\left (
  \begin{array}{cc}
   \xi & \Delta 
\\
   \Delta & - \xi 
  \end{array}
\right)
=
\xi \sigma_z
+
\Delta 
\sigma_x.
\end{eqnarray}
Thus, the impact of the complex nature of the order parameter is successfully eliminated.
Subsequently, we rotate along the $S_2$ axis as
\begin{eqnarray}
&&\mathcal{D}_y^{\dagger}(\theta_{\rm L})
\mathcal{D}_z^{\dagger}(\phi)
H_{\rm L} 
\mathcal{D}_z(\phi)
\mathcal{D}_y(\theta_{\rm L})
\nonumber \\
&&=
\sqrt{\xi^2+\Delta^2}
\left (
  \begin{array}{cc}
1  &  
0\\
0  &  
-1  \end{array}
\right)
\nonumber \\
&&=
\sqrt{\xi^2+\Delta^2}
\sigma_z
\end{eqnarray}
to obtain the diagonalised Hamiltonian.
The corresponding eigenstate for the positive energy becomes
\begin{eqnarray}
u_{\uparrow}
&=&
\mathcal{D}_z(\phi)
\mathcal{D}_y(\theta_{\rm L})
\left (
  \begin{array}{c}
1  \\
0  
\end{array}
\right) \\
&=&
{\rm e}^{-i \frac{\phi}{2}}
\left (
  \begin{array}{c}
 u_{\rm L}  \\
{\rm e}^{i \phi} v_{\rm L}  
\end{array}
\right)
=
\left (
  \begin{array}{cc}
    {\rm e}^{-i\frac{\phi}{2}} \cos \left( \frac{\theta_{\rm L}}{2} \right)    \\
    {\rm e}^{+i\frac{\phi}{2}}  \sin \left( \frac{\theta_{\rm L}}{2} \right)  
  \end{array}
\right),
\end{eqnarray}
which is the state for the rotated $|{\rm L} \rangle$.
The eigenstate for the negative energy becomes
\begin{eqnarray}
u_{\downarrow}
&=&
\mathcal{D}_z(\phi)
\mathcal{D}_y(\theta_{\rm L})
\left (
  \begin{array}{c}
0  \\
1  
\end{array}
\right)
\\
&=&
{\rm e}^{+i \frac{\phi}{2}}
\left (
  \begin{array}{c}
 -{\rm e}^{-i \phi} v_{\rm L}  \\
 u_{\rm L}  
\end{array}
\right)
=
\left (
  \begin{array}{cc}
   - {\rm e}^{-i\frac{\phi}{2}} \sin \left( \frac{\theta_{\rm L}}{2} \right)    \\
     {\rm e}^{+i\frac{\phi}{2}}  \cos \left( \frac{\theta_{\rm L}}{2} \right)  
  \end{array}
\right),
\end{eqnarray}
which also corresponds to $|{\rm R} \rangle$ for the standard time evolution, described by $H_{\rm R}$.
These states are orthogonal each other 
\begin{eqnarray}
u_{\uparrow}^{*}
u_{\downarrow}
&=&
\left (
  \begin{array}{cc}
    {\rm e}^{i\frac{\phi}{2}} \cos \left( \frac{\theta_{\rm L}}{2} \right)    &
    {\rm e}^{-i\frac{\phi}{2}}  \sin \left( \frac{\theta_{\rm L}}{2} \right)  
  \end{array}
\right)
\left (
  \begin{array}{c}
   - {\rm e}^{-i\frac{\phi}{2}} \sin \left( \frac{\theta_{\rm L}}{2} \right)    \\
     {\rm e}^{+i\frac{\phi}{2}}  \cos \left( \frac{\theta_{\rm L}}{2} \right)  
  \end{array}
\right)
\nonumber \\
&=&
0.
\end{eqnarray}

The obtained polarisation state is exactly the same form for the Bloch state in the 2-level system \cite{Saito20a}.
In fact, the polarisation state is described by quantum mechanical 2-level system with the $SU(2)$ symmetry \cite{Stokes51,Poincare92,Jones41,Yariv97,Baym69,Sakurai14,Saito20a}.
By making the superposition state of $|{\rm L} \rangle$ and $|{\rm R} \rangle$, we can construct any polarisation state.

\section{Discussions}

\subsection{Coherent macroscopic wavefunction}
Finally, we briefly describe the coherent many-body state of photons in a ray propagating along a GRIN fibre emitted from a laser source \cite{Yariv97,Saito20a,Saito20c}.
Here, we employ the chiral representation using the LR-basis, but we can also use other representations like the HV-basis \cite{Yariv97,Saito20a,Saito20c}.
The direction of the polarisation axis is not fixed solely from the Dirac equation.
However, in comparison with the electric field, obtained by using the corresponding coherent state, we can assign that the chiral axis of $S_3$ in the polarisation state is locked in to the direction of the propagation \cite{Saito20a}.
We use our convention that the left-circular polarised state is located at $S_3=1$ in the Stokes parameter notation, normalised to be unity of the magnitude \cite{Saito20a}. 
In this convention, the phase front of the electric field at the left-circular state is rotating towards the left, which is anti-clock-wise, seen from the detector side, using the standard right-handed $(x,y,z)$ coordinate \cite{Saito20a}.
Of course, any other convention is acceptable as far as the whole argument is closed consistently.

In the previous section, we confirmed that the dispersion relationship does not depend on the polarisation state in a GRIN fibre.
Therefore, the energy of photons are independent on the polarisation state, and photons can take any polarisation state, which is not determined solely from the fundamental Helmholtz equation or the Dirac equation.
This is a generic feature for all physical valuables, since a solution of a differential equation requires boundary conditions such as an initial condition for the time-evolution.
For the photon at the eigenstate, propagating in the GRIN fibre, there is no time-evolution for the polarisation state, such that the initial polarisation state will be maintained.
We discuss how the polarisation state is described in a many-body state of photons.

We consider an existence of single mode in a multi-mode fibre, but for the application of the multi-modes, we just need to consider the superposition of available modes, so that the extension is quite straightforward.
Photons are Bosons, such that multiple photons can occupy the same state.
If a macroscopic number of Bosons is occupying the same state, the macroscopic Bosons are described by a single wavefunction, and the corresponding macroscopic coherent assembly is called Bose-Einstein condensate \cite{Fetter03,Abrikosov75,Altland10,Nagaosa99}. 
Photons emitted from a laser source are also considered to be resided in such a Bose-Einstein condensed state, in which the phase coherence is expected, because the macroscopic number of photons are occupying the same state, described by one macroscopic wavefunction.
As we shall show, the wavefunction is nothing but the spinor wavefunction describing for the polarisation state of photons.
In order to maintain the phase coherence, we must allow the fluctuation of number of photons, since the number and the phase are conjugate each other.
This is achieved by considering the coherent states of photons \cite{Grynberg10,Fox06,Parker05}, 
\begin{eqnarray}
|\alpha_{\rm L} \rangle
&=&{\rm e}^{-\frac{|\alpha_{\rm L}|^2}{2}}
{\rm e}^{\alpha_{\rm L} \hat{a}_{\rm L}^{\dagger}}
|0\rangle 
\\
|\alpha_{\rm R} \rangle
&=&{\rm e}^{-\frac{|\alpha_{\rm R}|^2}{2}}
{\rm e}^{\alpha_{\rm R} \hat{a}_{\rm R}^{\dagger}}
|0\rangle , 
\end{eqnarray}
for left- and right-circular polarised states, respectively.
Here, we have defined the creation operator of $\hat{a}_{\rm \sigma}^{\dagger}$ and annihilation operator of $\hat{a}_{\rm \sigma}$, for the left and right polarisation states, $\sigma={\rm L}$ and $\sigma={\rm R}$, which satisfy the Bose commutation relationships $[\hat{a}_{\sigma},\hat{a}_{\sigma^{'}}] = 0$ and $[\hat{a}_{\sigma},\hat{a}_{\sigma^{'}}^{\dagger}] = \delta_{{\sigma},{\sigma}^{'}}$, using the Kronecker delta of $\delta_{{\sigma},{\sigma}^{'}}$ \cite{Sakurai67,Grynberg10,Fox06,Parker05,Saito20a,Saito20c}.
The coherent states \cite{Sakurai67,Grynberg10,Fox06,Parker05,Saito20a,Saito20c} are characterised by the complex number of $\alpha_{\sigma}$, which is an eigenvalue of the annihilation operator as
\begin{eqnarray}
\hat{a}_{\rm \sigma}
|\alpha_{\sigma} \rangle
=
\alpha_{\sigma}
|\alpha_{\sigma} \rangle,
\end{eqnarray}
whose conjugate becomes
\begin{eqnarray}
\langle \alpha_{\sigma} |
\hat{a}_{\rm \sigma}^{\dagger}
=
\langle \alpha_{\sigma} |
\alpha_{\sigma}^{*},
\end{eqnarray}
where $\alpha_{\sigma}^{*}$ is the complex conjugate of $\alpha_{\sigma}$.
Consequently, it is straightforward to obtain the quantum-mechanical average of the number operator
\begin{eqnarray}
\langle \alpha_{\sigma}|
\hat{a}_{\sigma}^{\dagger}
\hat{a}_{\sigma}
|\alpha_{\sigma} \rangle
&=&
|\alpha_{\sigma}|^2 , 
\end{eqnarray}
which is the number of photons, ${\mathcal N}_{\sigma}$, in the polarisation state of $\sigma$, and we have a sum rule for the total number of ${\mathcal N}={\mathcal N}_{\rm L}+{\mathcal N}_{\rm R}$.

Now, we are ready to assign the phases of the wavefunction to the polarisation state.
The polarisation state of a photon is characterised by the polar angle of $\theta$ and the azimuthal angle of $\phi$ in the Poncar\'e sphere (Fig. 5) \cite{Stokes51,Poincare92,Jones41,Yariv97,Baym69,Sakurai14,Saito20a}.
In the macroscopic coherent state, this could be implemented by assigning
\begin{eqnarray}
\alpha_{\rm L}
&=&
\sqrt{{\mathcal N}}
{\rm e}^{- i \frac{\phi}{2}}
\cos \left( \frac{\theta}{2} \right ) \\
\alpha_{\rm R}
&=&
\sqrt{{\mathcal N}}
{\rm e}^{+ i \frac{\phi}{2}}
\sin \left( \frac{\theta}{2} \right ) 
.
\end{eqnarray}
Consequently, we obtain the many-body wavefunction as 
\begin{eqnarray}
|\theta, \phi \rangle
&=&
|\alpha_{\rm L},\alpha_{\rm R}\rangle \\
&=&|\alpha_{\rm L}\rangle | \alpha_{\rm R}\rangle \\
&=&
{\rm e}^{-\frac{\mathcal N}{2}}
{\rm e}^{\alpha_{\rm L} \hat{a}_{\rm L}^{\dagger}}
{\rm e}^{\alpha_{\rm R} \hat{a}_{\rm R}^{\dagger}}
|0\rangle 
\\
&=&
{\rm e}^{-\frac{\mathcal N}{2}}
\exp 
\left [
		\sqrt{{\mathcal N}}
		{\rm e}^{- i \frac{\phi}{2}}\cos \left( \frac{\theta}{2} \right )
 		\hat{a}_{\rm L}^{\dagger}
\right ]
\nonumber \\
&& 
\exp 
\left [
		\sqrt{{\mathcal N}}
		{\rm e}^{+i \frac{\phi}{2}}\sin \left( \frac{\theta}{2} \right )
 		\hat{a}_{\rm R}^{\dagger}
\right ]
|0\rangle .
\end{eqnarray}
Please note that the entire wavefunction is characterised by single particle wavefunction with parameters $\theta$ and $\phi$.

We can confirm the spinor representation of the single particle wavefunction for the polarisation state in this many-body wavefunction, simply by applying the  complex electric field operator, 
\begin{eqnarray}
\bm{\hat{\mathcal{E}}}({\bf r},t)=
\sqrt{
  \frac{2 \hbar \omega_0}{\epsilon V}
  }
{\rm e}^{i kz -i \omega_0 t}
\left(
  \hat{a}_{\rm L}
  \hat{\bf l}
  +\hat{a}_{\rm R}
  \hat{\bf r}
\right),
 \end{eqnarray}
where $V=w_0^2 L_z$ with the length of $L_z$ for the fibre along $z$, $\hat{\bf l}=(\hat{\bf x}+i\hat{\bf y})/\sqrt{2}$ and $\hat{\bf r}=(\hat{\bf x}-i\hat{\bf y})/\sqrt{2}$ are complex unit vectors for directions of left and right polarisation states for the unit vectors of the coordinate of $\hat{\bf x}$ and $\hat{\bf y}$ along $x$ and $y$.
We obtain 
\begin{eqnarray}
\bm{\hat{\mathcal{E}}}({\bf r},t)
| \theta, \phi \rangle
&=&
\sqrt{
  \frac{2 \hbar \omega}{\epsilon V}
  }
{\Psi}({\bf r},t)
\left (
  \begin{array}{c}
    {\rm e}^{-i \frac{\phi}{2}}\cos \left( \frac{\theta}{2} \right)  \\
    {\rm e}^{+i \frac{\phi}{2}} \sin \left( \frac{\theta}{2} \right)  \ 
  \end{array}
\right)
| \theta, \phi \rangle,
\nonumber \\
 \end{eqnarray}
to confirm $| \theta, \phi \rangle$ is the eigenstate for $\bm{\hat{\mathcal{E}}}({\bf r},t)$, and the polarisation state is described by the Bloch state
\begin{eqnarray}
\langle \theta, \phi | {\rm Bloch} \rangle
&=&
\left (
  \begin{array}{c}
    {\rm e}^{-i \frac{\phi}{2}}\cos \left( \frac{\theta}{2} \right)  \\
    {\rm e}^{+i \frac{\phi}{2}} \sin \left( \frac{\theta}{2} \right)  \ 
  \end{array}
\right),
 \end{eqnarray}
which is indeed the solution of the Dirac equation for single photon.

\subsection{Stokes parameters}
We obtained the macroscopic quantum-mechanical wavefunction by the coherent state for photons. 
We can calculate relevant observables such as electric-field, magnetic-field, and so on as expectation values of this wavefunction \cite{Saito20a,Saito20c}.
Here, we show that the spin expectation values are Stokes parameters \cite{Saito20a,Saito20c} in Poincar\'e sphere (Fig. 5) by using $|\theta, \phi \rangle$.

Many-body spin operators \cite{Saito20a,Saito20c} for photons in LR-basis are defined as
\begin{eqnarray}
\hat{S}_{x}
&=&
\hbar 
     \bm{\hat{\psi}}_{\rm LR}^{\dagger}
\sigma_1
\bm{\hat{\psi}}_{\rm LR},
\\
\hat{S}_{y}
&=&
\hbar 
     \bm{\hat{\psi}}_{\rm LR}^{\dagger}
\sigma_2
\bm{\hat{\psi}}_{\rm LR},
\\
\hat{S}_{z}
&=&
\hbar 
     \bm{\hat{\psi}}_{\rm LR}^{\dagger}
\sigma_3
\bm{\hat{\psi}}_{\rm LR},
\end{eqnarray}
the spinor representation of the creation and annihilation field operators are defined as
\begin{eqnarray}
\bm{\hat{\psi}}_{\rm LR}^{\dagger}
&=&
( \hat{a}_{\rm L}^{\dagger}, \hat{a}_{\rm R}^{\dagger})
\\
\bm{\hat{\psi}}_{\rm LR}
&=&
\left(
  \begin{array}{c}
     \hat{a}_{\rm L} \\
     \hat{a}_{\rm R}
  \end{array}
\right) .
\end{eqnarray}
The spin operators are also re-written as
\begin{eqnarray}
\hat{S}_{x}
&=&
\hbar
\left(
\hat{a}_{\rm L}^{\dagger} \hat{a}_{\rm R}
+
\hat{a}_{\rm R}^{\dagger} \hat{a}_{\rm L}
\right) \\
\hat{S}_{y}
&=&
\hbar
\left(
-i
\hat{a}_{\rm L}^{\dagger} \hat{a}_{\rm R}
+i
\hat{a}_{\rm R}^{\dagger} \hat{a}_{\rm L}
\right)
\\
\hat{S}_{z}
&=&
\hbar
\left(
\hat{n}_{\rm L}
-
\hat{n}_{\rm R}
\right) .
\end{eqnarray}
Therefore, it is straightforward to calculate the quantum-mechanical expectation value as
\begin{eqnarray}
\langle \hat{S}_{x} \rangle
&=&
\hbar
\left(
{\alpha}_{\rm L}^{*} {\alpha}_{\rm R}
+
{\alpha}_{\rm R}^{*} {\alpha}_{\rm L}
\right) \\
&=&
\hbar {\mathcal N}
\left(
{\rm e}^{i \phi} \cos \left( \frac{\theta}{2} \right) \sin \left( \frac{\theta}{2} \right)
\right . \nonumber \\
&&
\left .
+
{\rm e}^{-i \phi} \cos \left( \frac{\theta}{2} \right) \sin \left( \frac{\theta}{2} \right)
\right) \nonumber \\
&=&
\hbar {\mathcal N}
\cos{\phi} \sin \theta
\\
\langle \hat{S}_{y} \rangle
&=&
\hbar
\left(
-i
{\alpha}_{\rm L}^{*} {\alpha}_{\rm R}
+i
{\alpha}_{\rm R}^{*} {\alpha}_{\rm L}
\right) \\
&=&
\hbar {\mathcal N}
\left(
-i {\rm e}^{i \phi} \cos \left( \frac{\theta}{2} \right) \sin \left( \frac{\theta}{2} \right)
\right . \nonumber \\
&&
\left .
+
i {\rm e}^{-i \phi} \cos \left( \frac{\theta}{2} \right) \sin \left( \frac{\theta}{2} \right)
\right) \nonumber \\
&=&
\hbar {\mathcal N}
\sin{\phi} \sin \theta
\\
\langle \hat{S}_{z} \rangle
&=&
\hbar
\left(
{\mathcal N}_{\rm L}
-
{\mathcal N}_{\rm R}
\right) \\
&=&
\hbar {\mathcal N}
\left(
\cos^2 \left( \frac{\theta}{2} \right)
-
\sin^2 \left( \frac{\theta}{2} \right)
\right) \nonumber \\
&=&
\hbar {\mathcal N}
\cos \theta .
\end{eqnarray}

For the total number of photons, $S_0$ is defined as
\begin{eqnarray}
\hat{S}_0
&=&
\hbar 
{\bm \psi}_{\rm LR}^{\dagger}
{\bf 1}
{\bm \psi}_{\rm LR} \\
&=&
\hat{a}_{\rm L}^{\dagger} \hat{a}_{\rm L}
+
\hat{a}_{\rm R}^{\dagger} \hat{a}_{\rm R},
\end{eqnarray}
and its expectation value for the coherent state becomes
\begin{eqnarray}
\hat{S}_0
&=&
\hbar {\mathcal N}.
\end{eqnarray}

We define the total spin vector operator as $\bm{\hat{\mathcal S}}=(\hat{S}_0,\hat{S}_x,\hat{S}_y,\hat{S}_z)$, and its expectation value by the coherent state is given by Stokes parameters \cite{Stokes51,Poincare92,Jones41,Yariv97,Goldstein11,Gil16,Pedrotti07,Hecht17,Payne52,Fano54,Collett70,Delbourgo77,Luis02,Luis07,Bjork10,Saito20a,Saito20c} 
\begin{eqnarray}
\bm{\mathcal S}
&=&
\langle \bm{\hat{\mathcal S}} \rangle 
=
\left (
  \begin{array}{c}
S_0 \\
S_1 \\
S_2 \\
S_3 
  \end{array}
\right )
=\hbar {\mathcal N}
\left (
  \begin{array}{c}
1 \\
\sin \theta \cos \phi \\
\sin \theta \sin \phi \\
\cos \theta \\
  \end{array}
\right ).
\end{eqnarray}

More generally, a ray in the fibre could contain an contribution from the incoherent state \cite{Stokes51,Poincare92,Yariv97,Goldstein11,Gil16}.
The phase degrees of freedom described by phases of $\theta$ and $\phi$ are responsible for the coherent parts, which are included in $S_1$, $S_2$, and $S_3$, while the total number of photons of ${\mathcal N}$ is responsible for both contributions, included in $S_0$.
Therefore, we can define the degree of polarisation to measure the coherence as a measure of polarisation, 
\begin{eqnarray}
d_{\rm p}=\frac{\sqrt{S_1^2+S_2^2+S_3^2}}{S_0},
\end{eqnarray}
which takes the value from 0 (unpolarised) to 1 (fully polarised).
Similarly, we can also define the degree of unpolarised light as 
\begin{eqnarray}
d_{\rm u}=\frac{S_0-\sqrt{S_1^2+S_2^2+S_3^2}}{S_0}=1-d_{\rm p},
\end{eqnarray}
which takes the value from 0 (fully polarised) to 1 (unpolarised).
The amount of polarised light ($d_{\rm p}$) plays a role of an order parameter to understand the amount of photons in a coherent state.

\begin{table*}[t]
\caption{\label{Table-VI}
Comparison of macroscopic coherent states in a laser source and superconductivity.
$SU(2)$ symmetry of polarisation state is broken upon lasing, described by Stokes parameters as expectation values of spin of photons.
Coherent state of photons is described by the Bose-Einstein condensation of photons to occupy the single mode described by a spinor wavefunction with fixed polarisation state.
The theory of superconductivity is explained by the Bose-Einstein condensation of Cooper pairs to zero momentum state for the centre-of-gravity motion.
}
\begin{ruledtabular}
\begin{tabular}{ccccc}
System&Broken symmetries & Phases & Order parameters & Wavefunction\\
\colrule
Laser&$SU(2)$&$\theta$ and $\phi$ & 
$\bm{\mathcal S}=\langle \bm{\hat{\mathcal S}} \rangle $ & 
$
\exp 
\left [
		\sqrt{{\mathcal N}}
		{\rm e}^{- i \frac{\phi}{2}}\cos \left( \frac{\theta}{2} \right )
 		\hat{a}_{\rm L}^{\dagger}
\right ]
\exp 
\left [
		\sqrt{{\mathcal N}}
		{\rm e}^{+i \frac{\phi}{2}}\sin \left( \frac{\theta}{2} \right )
 		\hat{a}_{\rm R}^{\dagger}
\right ]
|0\rangle 
$
 \\
Superconductivity & $U(1)$ & $\phi$ & $\Delta$ & 
$\prod_{{\bf k}} 
\exp
\left [
{\rm e}^{i \phi}
\tan \left ( \frac{\theta_{\bf k}}{2} \right )
\hat{b}_{{\bf  k}}^{\dagger}
\right ]|0 \rangle$
\end{tabular}
\end{ruledtabular}
\end{table*}

\subsection{Broken symmetry}
As we have seen above, the coherent state of photons, emitted from a laser source, is considered to be characterised by the broken $SU(2)$ state with fixed phases of $\theta$ and $\phi$.
In the original Helmholtz equation and the Dirac equation for photons, we have confirmed the rotational symmetry, expected for the GRIN fibre, considered in this paper. 
Nevertheless, in the coherent state, the superposition state with different number of photons are allowed, and the phases are fixed as a result of the Bose-Einstein condensation.
We have developed a fundamental theory to understand the quantum-mechanical origin of spin for photons in close analogy with the theory of superconductivity \cite{Bardeen57,Anderson58,Schrieffer71,Abrikosov75,Nambu59,Bogoljubov58,Saito19}.
Here, we discuss the similarity and the difference in more detail.

The most important difference is coming from the statistics of the elementary particles involved. 
Obviously, superconductivity is driven by electrons, which are Fermions \cite{Bardeen57,Schrieffer71}, and therefore, it is hard to expect the similarity to the Bose-Einstein condensation.
Nevertheless, the weak attractive interaction mediated by phonons or spin fluctuations could lead the formation of a Cooper pair, which is made by an electron with the wavevector ${\bf k}$ and spin up ($\uparrow$) and its time-reversal symmetric counter part of $-{\bf k}$ and spin down ($\downarrow$) electron.
The entire system is described by the BCS variational wavefunction
\begin{eqnarray}
|\phi \rangle=&
\prod_{{\bf k}} \left [
u_{{\bf k}}+
v_{{\bf k}}
\hat{c}_{{\bf  k} {\rm \uparrow}}^{\dagger}
\hat{c}_{{\bf -k} {\rm \downarrow}}^{\dagger}
\right ]|0 \rangle,
\end{eqnarray}
where $\hat{c}_{{\bf  k} {\rm \uparrow}}^{\dagger}$ and $\hat{c}_{{\bf -k} {\rm \downarrow}}^{\dagger}$ are the creation operator for an electron with spin up and down, respectively, and $u_{{\bf k}}$ and $v_{{\bf k}}$ are variational parameters
\begin{eqnarray}
u_{\bf k} &=& \cos \left ( \frac{\theta_{\bf k}}{2} \right ) \\
v_{\bf k} &=& {\rm e}^{i \phi} \sin \left ( \frac{\theta_{\bf k}}{2} \right ), 
\end{eqnarray}
to be defined by minimising the total free energy of the system to explain the superconducting phase transition \cite{Bardeen57,Schrieffer71}.
Here, what is important is the fact that all Cooper pairs are occupying the state with zero momentum, $\hbar {\bf k} - \hbar {\bf k}=0$.
In other words, the Cooper pairs, made of 2 Fermions for each, are occupying the same state with zero momentum among many other choices with arbitrary finite momentum.
In that respect, the Cooper pairs are exhibiting the Bose-Einstein condensation.
Moreover, the $U(1)$ phase degree of freedom is spontaneously broken, which is evident by the fixed phase of $\phi$.
Here, it is very important that $\phi$ has no dependence on ${\bf k}$ and the phase of $\phi$  is common for all Cooper pairs.
The phase coherence is allowed, because of the fluctuation of number of Cooper pairs, since the BCS state is allowing the superposition state of different number of pairs.

In order to see the similarity to our wavefunction of $|\theta, \phi \rangle$ for photons, we define the creation operator of Cooper pair, $\hat{b}_{{\bf  k}}^{\dagger}=\hat{c}_{{\bf  k} {\rm \uparrow}}^{\dagger} \hat{c}_{{\bf -k} {\rm \downarrow}}^{\dagger}$, and re-write the BCS wavefunction as 
\begin{eqnarray}
|\phi \rangle
&=&
\prod_{{\bf k}} 
u_{{\bf k}}
\left [
1
+
\frac{v_{{\bf k}}}{u_{{\bf k}}}
\hat{c}_{{\bf  k} {\rm \uparrow}}^{\dagger}
\hat{c}_{{\bf -k} {\rm \downarrow}}^{\dagger}
\right ]|0 \rangle \\
& \propto &
\prod_{{\bf k}} 
\left [
1
+
{\rm e}^{i \phi}
\tan \left ( \frac{\theta_{\bf k}}{2} \right )
\hat{b}_{{\bf  k}}^{\dagger}
\right ]|0 \rangle 
\\
& = &
\prod_{{\bf k}} 
\exp
\left [
{\rm e}^{i \phi}
\tan \left ( \frac{\theta_{\bf k}}{2} \right )
\hat{b}_{{\bf  k}}^{\dagger}
\right ]|0 \rangle ,
\end{eqnarray}
since $\hat{b}_{{\bf  k}}^n=0$ for all integers of $n \ge 2$.
The comparison of the theory of superconductivity with the $SU(2)$ theory for photons described in this paper is summarised in Table \ref{Table-VI}.
Here, we have just discussed a simple superconducting order parameter of $\Delta$ with $s$-wave symmetry, but it can contain internal structures, characterised by orbitals like $p$- and $d$-waves.
The same is true for our photonic state, and the confinement order parameter of $\Delta_n^m$ is indeed characterised by the orbital angular momentum of $m$ and the radial quantum number of $n$.

\section{Conclusion}
Spin is an inherent degree of freedom to characterise the polarisation state for a photon.
We have discussed the wavefunction for a photon described by a Helmholtz equation in a graded index fibre, which can be solved exactly using special functions such as Laguerre-Gauss and Hermite-Gauss functions.
The energy spectrum is described by the massive Schr\"odinger equation, which could be transferred to Klein-Gordon equation by a simple unitary transformation for shifting the energy due to the condensation for lasing, accompanied by the confinement in the fibre.
The two-dimensional Klein-Gordon equation is factorised to be the Dirac equation for a photon for allowing the quantum-mechanical probabilistic interpretation. 
The decoupled Dirac equation is exactly the same form with the spin equation of motion under the effective magnetic field.
We have shown the full spherical symmetry of the spin state, by the rotational symmetry of spin using $SU(2)$ operators.
The theory of spin state for a photon has a close similarity with the theory of superconductivity due to the intrinsic 2-level nature of polarisation state and paring state.
We have shown that the macroscopic quantum coherent state is a manifestation of the Bose-Einstein nature of the coherent state described by the broken $SU(2)$ symmetry of the polarisation state.
Consequently, the spin expectation values calculated by the coherent state becomes the Stokes parameters in the Poincar\'e sphere.

\section*{Acknowledgements}
This work is supported by JSPS KAKENHI Grant Number JP 18K19958.
The author would like to express sincere thanks to Prof I. Tomita for continuous discussions and encouragements.

\bibliography{Dirac}% Produces the bibliography via BibTeX.

\end{document}